# *Realistic model for radiation-matter interaction*

Richard A. Pakula
Science Applications International Corporation (SAIC), 4001 North Fairfax Dr., Suite 450, Arlington, VA 22203

## ABSTRACT

This paper presents a realistic model that describes radiation-matter interactions. This is achieved by a generalization of first quantization, where the Maxwell equations are interpreted as the electromagnetic component of the Schrödinger equation. This picture is complemented by the consideration of electrons and photons as real particles in three-dimensional space, following guiding conditions derived from the particle-wave-functions to which they are associated. The guiding condition for the electron is taken from Bohmian mechanics, while the photon velocity is defined as the ratio between the Poynting vector and the electromagnetic energy density. The case of many particles is considered, taking into account their statistical properties. The formalism is applied to a two level system, providing a satisfactory description for spontaneous emission, Lamb shift, scattering, absorption, dispersion, resonance fluorescence and vacuum fields. This model adequately describes quantum jumps by the entanglement between the photon and the atomic system and it will prove to be very useful in the simulation of quantum devices for quantum computers and quantum information systems. A possible relativistic generalization is presented, together with its relationship to QED.

03.65.-w, 14.70.Bh, 42.50.Ct, 42.50.-p

# Contents

# I. INTRODUCTION

A more intuitive understanding of the natural laws described by quantum mechanics is required to build more refined visualization and simulation tools for the development of quantum devices. Such an intuitive, realistic and non-local theory already exists for electrons and other material particles, as originally conceived by de Broglie [1, 2] and redefined by Bohm [3]. However, to the best of our knowledge a similar model does not exist for photons. This paper proposes a first approach towards the construction of such a theory. We notice that the first step towards the development of an intuitively friendly theory for photons is the construction of a first quantized version of quantum electrodynamics in space. The second step is to verify that one can indeed resolve the measurement problem and describe the quantum jumps in terms of real photon particles. The correct formulation of these two tasks will prove to be of precious value at the time of design and simulation of quantum computers and communication systems.

One can start by noting that the roots of quantum mechanics can be traced to an attempt for unification of the classical description of dispersion of light with the non-classical description provided by Bohr's atomic theory. In fact, both theories had been extensively verified experimentally, but seemed completely incompatible with one another. Dispersion theory implied a dynamical continuous atomic polarization, while Bohr's model implied the presence of stationary states separated by quantum jumps associated with the absorption and emission of electromagnetic radiation [4, 5]. The link was achieved by replacing the deterministic dynamical coefficients appearing in the polarization from the classical dispersion formula with the probabilistic Einstein coefficients from Bohr's atomic theory. That attempt, built exclusively on the consideration of the atomic dynamics, lead to the Kramers-Heisenberg dispersion formula [6] and eventually to the creation of the highly counterintuitive matricial form of quantum mechanics and to the introduction of noncommutative operators replacing dynamical variables in the classical equations of motion [7].

A realistic model is proposed here, which can reconcile classical dispersion theory and Bohr's atomic model. This model is based on the original proposal of Slater [5, 8], inspired in the de Broglie-Einstein ideas about the electromagnetic fields and photon particles. In Slater's model both, electromagnetic waves and photon particles were assigned physical existence in space, the waves, or photon-wave-functions, serving as guiding fields for the photons. One can verify that such a model is able to reconciliate wave and particle aspects of the interaction of radiation and matter, when the probability for absorption for an incident photon is made equal to the relative absorption of energy provided by the classical dispersion theory. This will be achieved as soon as photons are made to follow the Poynting vector, and its probability density made proportional to the electromagnetic energy density of the incident field. In that case, the same cross section obtained from the classical dispersion theory for the absorption of energy, would apply equally well to the probability for the absorption of a photon in Bohr's model. As a corollary, the transition amplitudes appearing in the Bohr model can be considered as numerically equal to the terms appearing in the classical dispersion coefficients as proposed by Ladenburg and followed by Kramers and Heisenberg [4, 5], in the steps previous to the creation of quantum mechanics.

While the interpretation of absorption as a virtual intermediate state in the process of dispersion in traditional quantum mechanics, induces one to consider both as two aspects of the same phenomenon, they correspond in our model to two different physical processes. Dispersion occurs every time an electromagnetic wave or photon-wave-function impinges on an atomic system, independently from the position of the real photon particle. It is responsible among others, for the generation of the index of refraction of transparent media. Absorption on the other hand, takes place only when one incident photon particle occurs to come to a distance closer than the absorption cross section and gets trapped by the atom, which now has been promoted to an excited state.

This paper develops a formal basis for these ideas towards a wider physical picture and is open for future improvement. It presents in a systematic and clear way the physical ideas implied by the use of photon-wave-functions and photon particles. To simplify the exposition in most of the paper, no consideration is given to any effect associated with the spin of the particles, or with relativistic dynamics. In Section VIII however, a relativistic generalization is presented and the relationship of this model with quantum electrodynamics (QED) is given.

Three different aspects of the interaction between light and matter provide the basis for our model. The first aspect is the transformation over time suffered by the intuitive description of quantum electromagnetic fields. This description started as a mathematical tool deprived from any real physical existence, and evolved to a form more reminiscent of classical electrodynamics in recent times. This evolution encouraged our visualization of the fields as having a reality in space, and expressed by the guiding conditions to which they give rise in this model. The second aspect is the development of a first quantization scheme for electrodynamics. In this scheme, the Maxwell equations are interpreted as the equivalent to the Schrödinger equation, and the electromagnetic fields are considered the photon-wave-functions. This is an important point, because in Bohmian mechanics, the guiding conditions are provided by the wave functions associated with the material particles in a first quantization theory. The third aspect is still an open question in quantum mechanics: can we explain quantum jumps?

This paper is organized as follows. In Section II, a review of the concepts leading to our model is provided. In Section III, the interpretation of the Maxwell equations as the single photon Schrödinger equation, the problems associated to the divergence of the fields at the sources, and a possible definition of attached and free fields are considered. In Section IV, the formalism associated with the generalized Schrödinger-Maxwell equation is presented, where photons are treated on the same footing as any other material particle. In Section V, the generalization to many particles and the selection of independent variables are analyzed. In Section VI, the guidance condition for the electron and photon in the extended configuration space is provided. In Section VII, the new formalism is applied to scattering, absorption, dispersion, resonance fluorescence, quantum jumps and vacuum fields in a two level system. In Section VIII, a possible relativistic generalization is presented, together with its relationship to QED. In Section IX, previous criticism raised again realistic models is considered, and in Section X, we present our conclusions.

## II. REVIEW OF THE CONCEPTS LEADING TO THIS MODEL

This paper proposes a reformulation of some of the ideas and equations prevailing in the description of quantum optics today. Therefore, a relatively complete review of the most relevant developments leading to the ideas exposed in this work is needed in order to gain the more general perspective required for a better appreciation of this proposal.

### A. Visualization of the quantum electromagnetic field.

We start considering how the intuitive visualization of the process of emission of electromagnetic radiation evolved with time since the first works at the beginning of quantum mechanics. In the framework of the old quantum theory of Bohr [9], emission of radiation was accomplished through a quantum jump from an atomic excited state to a lower energy state. In the theory of Dirac [10], this process was described by the abstract creation of a one-photon state of the electromagnetic field with well-defined momentum, represented by a set of quantum mechanical operators. These operators as well as the electron-wave-functions had no real existence in space. They were seen as a purely mathematical and formal representation of the process that was being described. An important property of that theory was the fact that it made no allowance for the radiation reaction fields, responsible for the decay of the oscillatory motion in classical electrodynamics. By contrast, quantum electrodynamics considered the emission process as induced by the so-called vacuum fields or vacuum fluctuations [11, 12].

In the 1960s, a description of spontaneous emission in terms of radiation reaction was developed, first as a semiclassical theory [13], and later making full use of the quantum mechanical equations of motion in the Heisenberg's picture [14]. In addition, the van der Waals and Casimir forces were recalculated and explained without the need for vacuum fluctuations [15].

In the 1970s the coherence properties of fields emitted by atoms were further analyzed [16], and it was shown that retardation effects, of long known history in classical electromagnetism could be invoked for the high frequency cut off in the non-relativistic calculation for the Lamb shift [17]. Retardation effects were also considered for the interaction between different atoms [18] in the cooperative effects described by Dicke's model [19].

In the early 1980s, working on causality effects in the proximity of atoms and molecules, Bykov [20] was able to show that the near fields operators follow exactly the same equations and have the same solutions as the classical fields. At the same time, Barut [21] started a series of papers showing that many of the effects previously considered of purely quantum origin, produced by the vacuum fields, could also be explained by classical fields, allowing for a more complicated mathematics. Among these effects, we can mention spontaneous emission and Lamb shift, van der Waals forces, Casimir forces, accelerating thermal fields, etc. In the following years, the works by the groups of Persico [22] and Power [23] yielded an increased visualization of the electromagnetic fields surrounding atoms. They also showed that causality could be fully re-established by requiring the retention of non-resonant terms in the fields, commonly through away in the rotating wave approximation [24].

All these works provided a more complete understanding of the electromagnetic fields surrounding the atoms and their interactions. As a result, the originally abstract and indeterministic picture of emission of radiation into a state of well-defined momentum was giving place to a more intuitive and 'realistic' picture in space and time, increasingly reminiscent of classical electrodynamics.

### B. First quantization schemes for the electromagnetic field.

Classical electrodynamics [25] and optics [26] show that electromagnetic fields display a very rich geometry, depending on the shape of the source currents and boundary conditions. The classical theory for the scattering by small particles [27, 28] makes use of spherical Bessel functions based on the work of Mie [29]. Also, classical antenna theory is concerned mostly with the near and far fields in the dipole approximation [30]. Similar treatments can be found in the semiclassical descriptions commonly used in X-ray diffraction and nuclear physics. X-ray diffraction by crystals can be explained in terms of the far field dipole emission or plane waves [31, 32, 33], as originally shown by Darwin [34] and by Ewald [35]. Nuclear radiation [36, 37, 38, 39, 40] has always been described in terms of the spherical harmonics developed by Hansen [41] and Heitler [42]. All these treatments contrast with the tradition in quantum optics, used to decompose the quantum fields in plane waves [43, 44, 45, 46, 47, 48], while working in the Coulomb gauge with transverse and longitudinal delta functions [49, 50, 51, 52].

A superficial consideration of quantum optics may create the impression that a geometrical distinction between classical and quantum fields really exist: while classical fields may have any geometry, quantum fields are restricted in most cases to plane waves. That this is really not the case is demonstrated by the source field expression in quantum optics [53, 54, 55], which shows that quantum electromagnetic fields produced by a point system have a spherical description, very similar to the classical one [56, 57, 58]. We find here again, that one could unify the rich geometric features of classical electrodynamics with the quantum properties of quantum optics, by considering a first quantized version of quantum electrodynamics, where the solutions to the Maxwell equations were considered wave functions for photon particles. This is not the standard way however, in which quantum electrodynamics has evolved. In fact, in semiclassical approximations [59], and in quantum optics [60], the systems are described in terms of particle positions and electromagnetic fields, while in second (or canonical) quantization [61, 62, 63, 64], the only physical quantities are the fields themselves. So, depending on the approach one selects, one has the choice to describe material particles by their position or by quantized fields, but not such option is offered for photons. The traditional theory provides no place for a photon position or velocity as a physical quantity in the description of the quantized electromagnetic field. We should remember however, that in the first approach to a quantum description of light, Einstein [65] proposed in 1905 that light was composed of real particles in space.

This state of facts had its origin in Dirac's original work [66] on the quantization of the electromagnetic field in 1927. In that paper, the electromagnetic field was not considered the equivalent to the wave function for photons, but Dirac considered instead the amplitude of the electromagnetic field itself as the physical quantity to be quantized. Each Fourier component of the field was considered equivalent to an independent harmonic oscillator coupled to the sources of radiation through the interaction Hamiltonian. This theory was further developed by Heisenberg and Pauli [67] and recapitulated by Fermi [68], reaching the general form that has been in use until today. The reason for considering the field as the primary variable to be quantized instead of the possible photon coordinates, seems to be the generalized reluctance to the idea that light could eventually be composed of material particles.

On the other hand, Landau and Peierls [69] published in 1930 a work about the generalized wave function of a system consisting of matter particles and of photon particles. This wave function consisted of a collection of functions with fix number of particles multiplied by time dependent coefficients in configuration space. They assumed that the role of photon-wave function was played exclusively by the electric field. There the photons were considered on an equal footing with the other particles. The main difference between the material particles and the photons was the existence of sources for the photon field. This fact was not considered harmful for the general development of the theory. The authors were able to show instead, that the sum of the norms of the functions belonging to different number of particles remained constant. The variation of the norm of these functions had two sources: the time variation of the respective coefficients, and the radiation sources for the photon fields. This was a marked difference with the customary methods of treating systems with variable number of particles, as described in the 1932 paper by Fock [70], where no references to photon sources was made at all. In that work, Fock demonstrated the full equivalence of working with the creation operator formalism of second quantization for bosons and fermions, and the use of wave functions with constant norm for different number of particles, but multiplied by time dependent coefficients.

The two ways of considering the quantization of the electromagnetic field (namely the Dirac and the Landau-Peierls formalism) were compared by Oppenheimer [71] in 1931, who concluded that Landau and Peierls source theory was not right because of the appearance of negative energies and photon densities. These were however, exactly the same problems initially attributed to Dirac's relativistic electron equation. Even when those problems didn't prevent a full success for Dirac's electron theory, no more work seems to have been done in this direction for photons until 1949, with the exception of a work by Kemmer [72] in 1939, who analyzed a spinorial expression for the Maxwell equations including the potentials in the Lorentz gauge. In 1949, Moliere [73] proposed the use of the electromagnetic field vectors in complex notation, previously developed by Silberstein in 1907 [74] as the photon-wave-function, and to use spherical solutions for the Maxwell's equations. Silberstein defined the complex electromagnetic fields functions as $\mathbf{F} = \mathbf{E} + ic\mathbf{B}$, where $\mathbf{E}$ and $\mathbf{B}$ were the electric and magnetic fields as usual. It can be noticed that Oppenheimer [71] in his previous paper also recognized the Silberstein functions as the appropriated wave functions for the photon.

For a period of about 20 years, work about photon-wave-functions focused mainly on the problem of localizability of particles with zero mass. The works by Pryce [75], Newton and Wigner [76], Whightman and collaborators [77] came invariably to the conclusion that zero mass particles with spin equal to one (like the photon) would not allow for the existence of wave functions able to localize a free particle in space as was possible with the other particles. As a corollary, they concluded that the photon as an independent particle could not exist. This was intimately related to the fact that the free electromagnetic field is divergence less in free space, and that a satisfactory and relativistically well-behaved position operator with commuting components could not be defined mathematically at that time [78]. We would like to remark that the problems presented in those papers were related to the impossibility for defining photon-wave-functions able to confine the photon to a precisely defined region of space. On the other hand, they could not prevent the possibility of postulating the existence of photons as particles in space independently from the properties of the wave functions to which they were associated. Two different solutions appeared to be possible at that time: 1) to renounce to the full localization of the photon, and accept only partial localization in small volumes or 2) propose that the photon mass could be different from zero and to work inside the framework of the Proca equation [79, 80] for spin 1 particles with nonzero mass. The first approach was proposed by Jauch and Piron [81], and Amrein [82] and developed explicitly in the model by Mandel [83], while the second approach was developed by Belinfante [84] and discussed by Bass and Schrödinger [85].

We should mention here the strong analogy that exists between the Silberstein's functions for the electromagnetic field and the solutions of the Dirac's equation for the relativistic electron. The electromagnetic fields are typically considered solutions of the Maxwell's second order differential equations, but in the Silberstein's notation, they are also solutions to a spinorial linear equation, as shown by Good [86] and Moses [87]. Similarly, the electron Dirac equation, which originally was presented in spinorial form, can also be rewritten as a set of second order differential equations for the electron function and for the time derivative of that function, as was shown by Feynman and Gell-Mann [88]. The original form of the Dirac equation and the spinorial form of the Maxwell equations are almost identical. Besides the finite electron mass, the only difference is that the Dirac's electron-wave-functions have four components, while the electromagnetic fields in complex notation have only three components, namely the three components of the electromagnetic field in space.

Surpassing the problems mentioned above with the particle interpretation for the photon, some works were published in the 1970-1990 period. For example the work by Moses [87], who showed that by working with spherical eigenvectors of the curl operator, not only interpretational advantages could be achieved, but also practical benefits, such as the rigorous determination of transition matrix elements. In the early 1980s, Cook [89] was able to show that the photon dynamics in free space was Lorentz covariant. After 1990, the main point was the realization that the wave function represented the probability for finding the photon energy at a given position in space. Among them, we can mention the works by I. Bialynicki-Birula [90,91], Inagaki [92], and Gersten [93]. The first one to consider the explicit interaction of radiation and matter in the context of photon-wave-functions appears to be Sipe [94], who also showed that the second quantization formalism, when applied to the photon-wave-functions, provides the same results as the traditional method in the description of the photon-atom interaction. Keller has developed a detailed model recently [95], where however, correlations between the matter fields and photon fields are not considered, and it is not possible to generate entanglement between the photons and the atomic system. A real breakthrough has been given recently by Margaret Hawton and coworkers [96], who obtained an explicit expression for the position operator with commuting components for the photon, therefore invalidating the work done in the decades of the '50 and '60 about the formal impossibility of defining photon as material particles.

### C. Quantum jumps and the photon particle.

From the beginning of the study of atomic processes, it seemed apparent that matter underwent discontinuous evolution. In the old quantum theory developed by Niels Bohr [9] in the 1910s, this was expressed by the assumption of the existence of 'quantum jumps'. In that theory, the atom could be found in the excited stationary states for a period comparable to the mean lifetime of the state, and then it decayed spontaneously, in a quantum jump, to the ground state, similar to the radioactive emission of particles by the atomic nucleus [97]. With the full development of quantum mechanics, these discontinuous transitions were later identified with the process of measurement. It was von Neumann [98] in 1932, who established clearly the Copenhagen interpretation for the act of measurement. In this interpretation, because of every measurement, the wave function collapses into the subspace of the Hilbert space spanned by the result of the measurement. Also as a result of the measurement, the observed system plus the measurement apparatus are left in a correlated state, indicating that if the apparatus shows a result α for the observable A, then the observed system is left in the subspace corresponding to the eigenvector of A with eigenvalue α. A following measurement of A will give inevitably the same value α. In this interpretation, it could be said that the act of measurement of the photon by the detector, collapses the state of the atom from the original excited state, to the ground state.

The exact definition of measured system and measuring apparatus is not a clear cut in quantum mechanics, but at the same time, different selections for the line separating system from measuring apparatus cannot modify the result of the measurement. In fact, one can argue that if we measure the photon at a given time, this means that the atom decayed to the ground state emitting the photon at a previous time allowing the photon to travel through space and reach the detector. Under this interpretation, it would be the photon the measuring apparatus, and the collapse of the atomic wave function would take place before the measurement by the detector, namely during the photon emission itself. A first experimental observation of these quantum jumps was performed in the late 1970s by the observation of the antibunching effect in the single atom emission by Mandel [99] and collaborators. This effect shows that after the emission of a photon by an atom, one has to wait a finite amount of time in order for the atom to be able to emit a second photon, even when the laser pump remains at a constant power. This fact is interpreted by saying that if the atom collapses (jumps) to the ground state after the emission of the first photon, it will need some time to be re-excited again and re-emit a second photon as shown by the quantum regression theorem [100].

Since then, quantum jumps have been observed in other experiments [101, 102] proposed initially by Dehmelt [103] in 1975, where the atoms get shelved in an excited state and are unable to absorb or scatter radiation from the ground state. After the first works directed to understand this phenomenon [104], numerous theoretical work has been done in order to incorporate the quantum jumps into the repertoire of physics [105]. On the theoretical side, the inclusion of quantum jumps has been used as an alternative approach in simulations to reproduce the time evolution of the density matrix. In these simulations, the average system is described as a collection of many individual wave functions following periods of continuous evolution and random quantum jumps followed by the renormalization of the wave function [106].

In spite of all these developments, the main question about the quantum jumps remains the same since the time of Bohr, namely: what triggers the jump? Traditional quantum mechanics is unable to answer this question, however realistic theories can provide an answer at the time they are solving the problem of measurement. Entanglement of the wave function, makes the quantum potential dependent on the position of all the particles composing the system. When one particle enters some specific wave-packet that has no spatial overlapping with the other ones, the only part of the wave function felt by the remaining particles in the system is the one directly correlated to the one that is occupied. A simple explanation offers by itself in the case of photon emission: if the photon leaving the atom is correlated only with the atomic ground state, after emission the electron sees only the ground state wave-function and stops oscillating. This should happen in a period comparable with the time it takes to the photon to leave the atom vicinity; we can say that a 'quantum jump' has occurred.

# III. PHOTON-WAVE FUNCTIONS

In this Section, we start with the development of our formalism. In part A we provide the present status about the consideration of the electromagnetic fields and potentials as the photon-wave-functions. In part B we consider the separation of the electromagnetic fields into free and attached to the particles. This separation will prove to be of fundamental importance for the generation of entanglement between the electromagnetic fields and the atomic systems.

### A. Maxwell's equations as the single photon Schrödinger equation

Induced by the developments described in the previous Section, we propose to consider the electromagnetic field as representing the wave function for photons. In support of this idea, we mention that Moses [87, 107] was able to show that Maxwell's equations take the form:

$$-\frac{1}{i}\frac{\partial}{\partial t}\Psi_{rad} = H_0\Psi_{rad} - 4\pi\Phi ,$$

where $\Psi_{rad}(\mathbf{r}_{ph})$ is a four-element column matrix constructed with the electric and magnetic fields, $\Phi(\mathbf{r}_{ph})$ is a source-current term, the Hamiltonian $H_0$ is given by $H_0 = \frac{1}{i}\boldsymbol{\alpha}\cdot\nabla$, and $\boldsymbol{\alpha} = (\alpha^x, \alpha^y, \alpha^z)$ is a set of 4x4 matrices. With these definitions, the similarity with the Dirac equation for the electron is evident, and in the absence of sources, their only difference is the existence of a finite electron mass. We can mention that conversely, the creation of an electron-positron pair by a gamma ray also admits a description allowing for a source term [108, 109].

When we consider the interaction between matter and light, we find that the Schrödinger equation involves the electromagnetic potentials, in addition to the electric and magnetic fields. We need therefore to find an expression where these potentials are available at the same time than the electric and magnetic fields. In some sense, the potentials store information about the history of the fields, and deserve to be considered, as noted by Bohm and Aharanov [110], on the same footing as the electromagnetic fields. In order to keep track of the fields as well as the potentials, Kemmer [72] developed in 1939 a spinorial representation of the Maxwell equations, including not only the electric and magnetic fields, but also the electromagnetic potentials in the Lorentz gauge. The time evolution of this spinor was governed by the following Schrödinger-like equation:

$$i\hbar \frac{\partial \Psi_{rad}(\mathbf{r}_{ph},t)}{\partial t} = H_{ph}\Psi_{rad}(\mathbf{r}_{ph},t) + J(\mathbf{r}_{ph},t) \tag{1}$$

where $J(\mathbf{r}_{ph},t)$ was a new source function. In Kemmer's approach, the Schrödinger-like equation represented the Maxwell's equations:

$$\frac{\partial \mathbf{E}}{\partial t} = \nabla \times \mathbf{B} - \mathbf{j}, \frac{\partial \mathbf{B}}{\partial t} = -\nabla \times \mathbf{E}, \frac{\partial \mathbf{A}}{\partial t} = -\mathbf{E} - \nabla \phi, \frac{\partial \phi}{\partial t} = \nabla \cdot \mathbf{A}$$

where the last equation is the Lorentz condition for the potentials. We remark here that eq. (1) holds as an alternative formulation of the classical Maxwell equations, and as the Heisenberg equation of motion fulfilled by the electromagnetic field operators in quantum electrodynamics. Kemmer also studied the relativistic invariance properties of the given matrix $H_{ph}$, but we are not going into the details here. We only mention that this representation is known as the Kemmer-Duffin-Harish-Chandra formalism [72]. Of course, one can rewrite Maxwell's equations following this formalism in the Coulomb gauge, at the price of losing Lorentz invariance.

### B. Separation into attached and free fields

It is well known that electromagnetic fields diverge at the position of the particles. Therefore, the electromagnetic fields are not normalizable and an infinite amount of energy is assigned to them. This is in distinction with the wave functions for electrons, which are square integrable over the full space, and allow for an easy manipulation inside the Hilbert space.

The purpose of this section is to show how it is possible to separate the electromagnetic field in two parts. One part attached to the particle will be the responsible for the divergences and non-normalizability of the electromagnetic fields. The other part, associated with the radiated energy, will be normalizable and incorporated into the first quantization formalism of the Hilbert space.

In relation to the question of energy content of the system, we can avoid the divergence problems of quantum electrodynamics by assigning the momentum and energy contents of the system to the particles, and not to the fields. The attached fields will play a role analogous to potential energies, and therefore can diverge to infinity in some regions, without implying an infinite energy for the physical system.

We will encounter situations where the number of particles present in our system changes with time. It is clear that the creation of a particle modifies the number of degrees of freedom of the system by adding to the position and velocities of the already existing particles, those corresponding to the newly created one. We can however reduce the use of a variable number of photons, by assuming that photons that have been coherently scattered still survived, confined inside the fields attached to the charged particles. This point of view is in line with the original approach followed by Dirac [66] where the vacuum state was just the ground state for the oscillators representing the radiation field. In that approach, the act of emitting a photon merely changed the state but not the number of degrees of freedom of the system.

Another advantage of considering attached fields is that they will allow us to generate entanglement between the different states of the radiation field and the different states of the atomic system.

In order to develop this idea further, let's consider the Maxwell's equation for the electric field:

$$\left(\nabla^2 - \frac{1}{c^2}\frac{\partial^2}{\partial t^2}\right)\mathbf{E}(\mathbf{r},t) = \mu_0 \frac{\partial \mathbf{J}(\mathbf{r},t)}{\partial t}$$

It is clear that this equation describes two subsystems, the particle represented by the current term, and the fields. Let's rewrite the electric field as the sum of two parts, which we might call the attached fields linked to the particle, and the free fields propagating in free space as $\mathbf{E}(\mathbf{r},t) = \mathbf{E}_{att}(\mathbf{r},t) + \mathbf{E}_{free}(\mathbf{r},t)$ with $\mathbf{E}_{att}(\mathbf{r},t)$ and $\mathbf{E}_{free}(\mathbf{r},t)$ the attached and free fields respectively. Now we can rewrite the Maxwell equation as follows:

$$\left(\nabla^2 - \frac{1}{c^2}\frac{\partial^2}{\partial t^2}\right)\left(\mathbf{E}_{att}(\mathbf{r},t) + \mathbf{E}_{free}(\mathbf{r},t)\right) = \mu_0 \frac{\partial \mathbf{J}(\mathbf{r},t)}{\partial t} .$$

Nobody can prevent us from defining an effective current density by the following expression:

$$\mu_0 \frac{\partial \mathbf{J}_{free}(\mathbf{r},t)}{\partial t} = \mu_0 \frac{\partial \mathbf{J}(\mathbf{r},t)}{\partial t} - \left(\nabla^2 - \frac{1}{c^2}\frac{\partial^2}{\partial t^2}\right)\mathbf{E}_{att}(\mathbf{r},t)$$

In terms of this effective current density, we can rewrite the Maxwell equation for the free fields as

$$\left(\nabla^2 - \frac{1}{c^2}\frac{\partial^2}{\partial t^2}\right)\mathbf{E}_{free}(\mathbf{r},t) = \mu_0 \frac{\partial \mathbf{J}_{free}(\mathbf{r},t)}{\partial t} .$$

One can reinterpret this equation, as describing again two subsystems, but now they are the effective current source for the free fields (which we can interpret as the original electron current together with the attached fields), and the free radiation fields.

This is not a new idea; in fact, different models have proposed different lines separating attached from radiated fields. A first example is just given by the Coulomb gauge, where the longitudinal electrostatic fields play the role of attached fields, and the transverse fields are considered radiated or free [50]. The source current for the free fields becomes the well-known transverse current. Kramers [111], Pauli and Fierz [112] included the magnetostatic fields derived from the stationary part of the velocity, in the attached fields. Indeed the idea of mass renormalization of quantum electrodynamics can be interpreted as a relativistic generalization of the Kramers transformation [113, 114, 115], where the energy associated with the magnetostatic attached fields is incorporated into the experimentally observed electron mass. Finally, Keller included a transverse attached field in the description of an oscillating dipole, and provided a unitary transformation leading to remove it [116]. For an oscillating dipole, we can see that all models draw the dividing line at a distance of about one wavelength from the source.

In this work, we will define the free fields as the acceleration fields, and the attached fields as the velocity fields. With this definition, we verify that the attached fields are dominant at a distance smaller than a wavelength from the atomic system, while the free fields are dominant at a distance larger that a wave length from the atomic center.

We anticipate here, that in the framework of a realistic model, photon particles should respond to both attached and radiated fields. As long as the photon is closer to the atom than one wavelength, the attached fields will govern its dynamics, so we can say that the photon belongs to the atomic system, it is an absorbed photon. As soon as the photon leaves the proximity of the atomic system, it will be governed by the radiated fields and become a free photon, an emitted photon. Later in Section VII, we will analyze the objections raised against the concept of a field attached to a particle.

According to our definition, the attached fields for charged point particles are given by [117]:

$$\mathbf{E}_{att}(\mathbf{r},t) = e \left[ \frac{\hat{\mathbf{n}} - \mathbf{v}/c}{\gamma^2 (1 - \hat{\mathbf{n}} \cdot \mathbf{v}/c)^3 r^2} \right]_{ret}, \quad \mathbf{B}_{att}(\mathbf{r},t) = \left[ \hat{\mathbf{n}} \times \mathbf{E}_{att}(\mathbf{r},t) \right]_{ret}, \quad (2)$$

and the free fields by

$$\mathbf{E}_{free}(\mathbf{r},t) = \frac{e}{c} \left[ \frac{\hat{\mathbf{n}} \times \{(\hat{\mathbf{n}} - \mathbf{v}/c) \times \dot{\mathbf{v}}/c\}}{(1 - \hat{\mathbf{n}} \cdot \mathbf{v}/c)^3 r} \right]_{ret}, \quad \mathbf{B}_{free}(\mathbf{r},t) = \left[ \hat{\mathbf{n}} \times \mathbf{E}_{free}(\mathbf{r},t) \right]_{ret} \quad (3)$$

where all the quantities and positions have to be taken a t a retarded time $t' = t - r/c$ and $r$ refers to the radius vector starting at the position of the particle and ending at the field position. It is very important to note that the free fields diverge at most as $1/r$ as we approach the particle:

$$\mathbf{E}_{free} \xrightarrow[r \to 0]{} \frac{1}{r} \quad \mathbf{B}_{free} \xrightarrow[r \to 0]{} \frac{1}{r}.$$

In this way, their norm and also their traditionally assigned energy content remain finite after integration over space. We can say that this partition into attached and free fields is relativistically invariant. The velocity fields transform among themselves, without mixing with the acceleration fields. Free and attached fields satisfy the equations:

$$\nabla \cdot \mathbf{E}_{free}(\mathbf{r},t) = 0$$
$$\nabla \cdot \mathbf{B}_{free}(\mathbf{r},t) = 0$$

$$\frac{1}{c^2}\dot{\mathbf{E}}_{free}(\mathbf{r},t) = -\vec{\nabla}\times\mathbf{B}_{free}(\mathbf{r},t) + \mathbf{j}^{E,free}(\mathbf{r},t)$$

$$\dot{\mathbf{B}}_{free}(\mathbf{r},t) = -\nabla\times\mathbf{E}_{free}(\mathbf{r},t)$$

and

$$\nabla\cdot\mathbf{E}_{att}(\mathbf{r},t) = 4\pi\rho$$

$$\nabla\cdot\mathbf{B}_{att}(\mathbf{r},t) = 0$$

$$\frac{1}{c^2}\dot{\mathbf{E}}_{att}(\mathbf{r},t) = -\vec{\nabla}\times\mathbf{B}_{att}(\mathbf{r},t) + \mathbf{j}^{E,att}(\mathbf{r},t)$$

$$\dot{\mathbf{B}}_{att}(\mathbf{r},t) = -\nabla\times\mathbf{E}_{att}(\mathbf{r},t)$$

Where we define

$$\mathbf{j}_i^{E,free}(\mathbf{r},t) = \sum_j \int \mathbf{j}_j^E(\mathbf{r}',t)\delta_{i,j}^{free}(\mathbf{r}-\mathbf{r}')d\mathbf{r}'$$

$$\mathbf{j}_i^{E,att}(\mathbf{r},t) = \sum_j \int \mathbf{j}_j^E(\mathbf{r}',t)\delta_{i,j}^{att}(\mathbf{r}-\mathbf{r}')d\mathbf{r}'$$

where $\delta_{i,j}^{att}(\mathbf{r}-\mathbf{r}') = \delta_{i,j}(\mathbf{r}-\mathbf{r}') - \delta_{i,j}^{free}(\mathbf{r}-\mathbf{r}')$, in a similar way to the definition of the transverse and longitudinal delta functions

These equations can be written in matricial form. The properties and symmetries of these matrices are similar to those considered before for the Kemmer-Duffin-Harish-Chandra formalism [72], the only difference being that the sources are associated with the attached and free fields respectively.

In the dipole approximation, the terms proportional to the first time derivative of the dipole and the second time derivative of the dipole correspond to the velocity and acceleration fields found in the non-relativistic approximation. The term proportional to the dipole corresponds to the deviation of the position of the charge from the origin of coordinates. Here $p(t)$ represents the magnitude of the dipole at time t, assumed to point in the z direction, and to be located at the origin of the system of coordinates. We find that the attached fields are:

$$\mathbf{E}_{d,att} = \left(\frac{p(t-r/c)}{r^3} + \frac{\dot{p}(t-r/c)}{cr^2}\right)\left(2\cos(\theta)\hat{\mathbf{r}} + \sin(\theta)\hat{\boldsymbol{\theta}}\right),$$

$$\mathbf{B}_{d,att} = \frac{\dot{p}(t-r/c)}{cr^2}\sin(\theta)\hat{\boldsymbol{\varphi}},$$

and the free fields are:

$$\mathbf{E}_{d,free} = \frac{\ddot{p}(t-r/c)\sin(\theta)\hat{\boldsymbol{\theta}}}{c^2 r}$$

$$\mathbf{B}_{d,free} = \frac{\ddot{p}(t-r/c)}{c^2 r}\sin(\theta)\hat{\boldsymbol{\varphi}}$$

In order to find the source term for the radiation fields in the dipole approximation, we notice that they are always perpendicular to the radius vector connecting the dipole with the test point. Therefore, what we need is just the projection of the fields and the source into the r-perpendicular direction. This is the same as the transverse delta function, but now in r-space, this is

$$\delta_{i,j}^{free}(\mathbf{r}) = \delta_{i,j}(\mathbf{r}) - \frac{r_i r_j}{r^2}.$$

## IV. THE SCHRÖDINGER-MAXWELL EQUATION

In this section, we propose a new scheme for the association of electrodynamics and quantum mechanics, by providing a combined Schrödinger-Maxwell equation, defined in a generalized configuration space. This space includes not only the traditional particles as electrons and protons, but also the photon particles as well. The generalized equation provides a unified first quantized theory for all those particles in space.

We start with the observation that in traditional many-particle quantum mechanics in first quantization, the Schrödinger equation reads

$$i\hbar \frac{\partial F(r_1, r_2, \ldots, r_n)}{\partial t} = \left[ \sum_i \hat{K}_i + \sum_{i,j} \hat{V}_{ij}(\vec{r}_i, \vec{r}_j) \right] F(r_1, r_2, \ldots, r_n) \quad (4)$$

where $F(r_1, r_2, \ldots, r_n)$ is the wave function defined in configuration space. The operator $\sum_i \hat{K}_i + \sum_{i,j} \hat{V}_{ij}(\vec{r}_i, \vec{r}_j)$ is the Hamiltonian where the first term $\sum_i \hat{K}_i = \sum_i \frac{1}{2} m_i \nabla_{r_i}^2$ represents the sum over the kinetic energy of the particles, and the second term $\sum_{i,j} \hat{V}_{ij}(\vec{r}_i, \vec{r}_j)$ the sum over all interaction energies. The wave function can be written, in general as the sum over products of the wave functions for individual particles:

$$F(r_1, r_2, \cdots r_n) = \sum_{states} b_{i,j\cdots k} \psi_i(\mathbf{r}_1, t) \psi_j(\mathbf{r}_2, t) \cdots \psi_k(\mathbf{r}_n, t) \quad (5)$$

Comparison of eq. (1) and eq. (4) induces us to consider the identification of the photon hamiltonian $H_{ph}$ from eq. (1) with the photon kinetic energy and $J(\mathbf{r}_{ph}, t)$ with the photon interaction energy. This allows us to write a generalized wave equation as:

$$i\hbar \frac{\partial F(r_1, r_2, \cdots, r_n, r_{ph})}{\partial t} = \hat{H} F(r_1, r_2, \cdots, r_n, r_{ph}) \quad (6)$$

where $F(r_1, r_2, \cdots, r_n, r_{ph})$ is the wave function defined in the generalized configuration space and we define $\hat{H}$ as the Schrödinger-Maxwell Hamiltonian

$$\hat{H} = \underbrace{\hat{K}(\mathbf{r}_e) + \hat{V}(\mathbf{r}_e)}_{Schrödinger} + \underbrace{\hat{H}_{ph}(\mathbf{r}_{ph}) + \hat{\mathbf{j}}(\mathbf{r}_{ph})}_{Maxwell}. \tag{7}$$

a sum of the Schrödinger and the Maxwell operators, this last one in the Kemmer-Duffin-Harish-Chandra formalism, where to simplify the notation, we have considered the interaction of a single photon with an atomic system. For the wave function we write in analogy with eq. (5) products of the form

$$F(r_1, r_2, \cdots, r_n, r_{ph}) = \sum_{states} b_{i,j,\cdots k,l,m,n} \psi_i(\mathbf{r}_1, t) \psi_j(\mathbf{r}_2, t) \cdots \psi_k(\mathbf{r}_n, t) \times$$
$$\times \Psi_l^{rad}(\mathbf{r}_{ph1}, t) \Psi_m^{rad}(\mathbf{r}_{ph2}, t) \cdots \Psi_n^{rad}(\mathbf{r}_{pho}, t) \tag{8}$$

where now $\psi_i(\mathbf{r}_1, t)\psi_j(\mathbf{r}_2, t) \cdots \psi_k(\mathbf{r}_n, t)$ represents the product of electron-wave-functions and $\Psi_l^{rad}(\mathbf{r}_{ph1}, t)\Psi_m^{rad}(\mathbf{r}_{ph2}, t) \cdots \Psi_n^{rad}(\mathbf{r}_{pho}, t)$ the product of photon-wave-functions. In support of this model, we can mention that the Heisenberg equation of motion for the electron-wave-function operator $\hat{\psi}(\mathbf{r}, t)$ in second quantization is identical with the Schrödinger equation for a single electron $\psi(\mathbf{r}, t)$. Also, the Heisenberg equation of motion for the electromagnetic field operator $\hat{\Psi}_{rad}(\mathbf{r}, t)$, in quantum electrodynamics, is the same as the Maxwell equations for the classical fields $\Psi_{rad}(\mathbf{r}, t)$ in the Kemmer formalism as given in eq. (1). In the same way, the equation of motion for the product of the operators $\hat{\psi}(\mathbf{r}, t)\hat{\Psi}_{rad}(\mathbf{r}, t)$ will be given by

$$i\hbar \frac{\partial(\hat{\psi}(\mathbf{r}, t)\hat{\Psi}_{rad}(\mathbf{r}, t))}{\partial t} = \hat{\Psi}_{rad}(\mathbf{r}, t) i\hbar \frac{\partial \hat{\psi}(\mathbf{r}, t)}{\partial t} + \hat{\psi}(\mathbf{r}, t) i\hbar \frac{\partial \hat{\Psi}_{rad}(\mathbf{r}, t)}{\partial t} =$$
$$= \hat{\Psi}_{rad}(\mathbf{r}, t)\hat{H}_{Schr}\hat{\psi}(\mathbf{r}, t) + \hat{\psi}(\mathbf{r}, t)\hat{H}_{ph}\hat{\Psi}_{rad}(\mathbf{r}, t) = (\hat{H}_{Schr} + \hat{H}_{ph})\hat{\psi}(\mathbf{r}, t)\hat{\Psi}_{rad}(\mathbf{r}, t) \tag{9}$$

where we have used the Heisenberg equation of motion for the operators $\hat{\psi}(\mathbf{r}, t)$ and $\hat{\Psi}_{rad}(\mathbf{r}, t)$ and the fact that they commute at equal times. We can see that the expression provided by eq. (9) is the same as eq. (6) and eq. (7).

The photon-wave-functions appearing in eq. (8) can be free or attached. We can find terms of the form $\Psi_l^{free}(\mathbf{r}_{phi}, t)\Psi_m^{att}(\mathbf{r}_{phi}, t)$, where both functions share the same photon coordinate. The rule we are going to use is that free photon-wave-functions will correspond to the existence of real photons in space, fulfilling energy requirements, and therefore multiplying electron states corresponding to the lower state of the transition originating the emission of radiation. Attached states will be considered as belonging to the real charged particle given rise to the fields. They will multiply both states corresponding to the transition taking place.

The solutions to equation (6) have their counterpart on the different states of radiation and matter in quantum optics. An important role in quantum optics is played by the so called vacuum states, namely the states of the radiation field with no photons present. We will introduce the equivalent ot vacuum states also in our description provided by eq. (8). They will be given by terms of the form $\Psi_{vac}^{free}(\mathbf{r}_{ph,null})$ and $\Psi_{vac}^{att}(\mathbf{r}_{ph,null})$, where the first one corresponds to emitted fields and the second one to attached fields. The difference will be that these functions are not associated with any real photon at all, therefore they have an argument provided by $\mathbf{r}_{ph,null}$. The functions $\Psi_{vac}^{free}(\mathbf{r}_{ph,null})$ may appear multiplying the upper state of energy conserving transitions, or both states from energy non-conserving transitions. The functions $\Psi_{vac}^{att}(\mathbf{r}_{ph,null})$ will multiply both states of energy non-conserving transitions. To exemplify the previous considerations, let's write down the wave function for the case of a single electron and a single photon in the process of emission of radiation:

$$F(\mathbf{r}_e,\mathbf{r}_{ph},t) = \left[a\psi_e(\mathbf{r}_e,t)\Psi_{vac}^{free}(\mathbf{r}_{ph,null}) + b\psi_g(\mathbf{r}_e,t)\Psi_{rad}^{free}(\mathbf{r}_{ph},t)\right]\Psi_{rad}^{att}(\mathbf{r}_{ph},t),$$

where $\psi_e(\mathbf{r}_e,t)$ and $\psi_g(\mathbf{r}_e,t)$ are the atomic excited and ground state respectively.

We will adopt for the fields the multipole expansion. We note that the adoption of the multipole formalism is equivalent to the description provided by the Power-Zienau-Wooley transformation [118, 119]. We adopt this description because it removes the static longitudinal fields that are distinguishable from the oscillating fields, while maintaining full causality. The static portion of the attached fields is automatically included in the Coulomb potential.

Neglecting terms quadratic in the vector potential, eq. (6) can be written in more detail as:

$$i\hbar\frac{\partial F}{\partial t} = \left(\hat{H}_{at} + \hat{H}_I + \hat{H}_{ph}^{free} + \hat{H}_{ph}^{att}\right)F$$

where $\hat{H}_{at}$ is the usual atomic Hamiltonian. The terms $\hat{H}_{ph}^{free}$ and $\hat{H}_{ph}^{att}$ represent matrices for free fields and attached fields, respectively. The interaction Hamiltonian $\hat{H}_I$ has two terms $\hat{H}_I = \hat{\mathbf{j}}_{rad} + \hat{H}_{I,at}$, the first term $\hat{\mathbf{j}}_{rad} = \hat{\mathbf{j}}_{rad}^{free} + \hat{\mathbf{j}}_{rad}^{att}$ gives rise to the source current appearing in the Maxwell's equations for the free and attached fields, and the second term $\hat{H}_{I,at}$ corresponds to the standard interaction term in the particle Hamiltonian discussed in many text books. These two terms connect wave functions with free fields to wave functions with and without free fields, these last one in a similar way as the theory by Landau and Peierls [69]. The explicit form for the source current will be discussed later in Section VI. For the second term, we propose the following form:

$$\hat{H}_{I,at} = \sum_i q_i \mathbf{p}_i \cdot \left(\hat{\mathbf{A}}(\mathbf{r}_{ph}) + \hat{\mathbf{A}}^\dagger(\mathbf{r}_{ph})\right)\delta(\mathbf{r}_i - \mathbf{r}_{ph}),$$

where $\hat{\mathbf{A}}(\mathbf{r}_{ph}) + \hat{\mathbf{A}}^{\dagger}(\mathbf{r}_{ph})$ represents the total (attached plus free) vector potential operators and $\mathbf{p}_i = \frac{\hbar}{i}\nabla_i$ is the usual momentum operator for the $i$-particle. The sum extends over all charged particles with coordinates $\mathbf{r}_i$, and $\mathbf{r}_{ph}$ is the coordinate in configuration space assigned to the photon particle, the Dirac's delta ensures that the field is taken at the position of the particle.

The photon-wave-functions, following the Kemmer-Duffin-Harish-Chandra formalism are given by $\Psi(\mathbf{r}_{ph},t) = \begin{pmatrix} \mathbf{E}(\mathbf{r}_{ph},t) \\ \mathbf{B}(\mathbf{r}_{ph},t) \\ \mathbf{A}(\mathbf{r}_{ph},t) \\ \phi(\mathbf{r}_{ph},t) \\ \mathbf{1} \\ \mathbf{1} \\ \mathbf{1} \\ 1 \end{pmatrix}$ and for the photon vacuum state we define

$\Psi_{vac}(\mathbf{r}_{ph,null}) = \begin{pmatrix} \mathbf{E}(\mathbf{r}_{ph,null},t) \\ \mathbf{B}(\mathbf{r}_{ph,null},t) \\ \mathbf{A}(\mathbf{r}_{ph,null},t) \\ \phi(\mathbf{r}_{ph,null},t) \\ \mathbf{1} \\ \mathbf{1} \\ \mathbf{1} \\ 1 \end{pmatrix}$. We assume that no real photon is ever associated with the coordinate $\mathbf{r}_{ph,null}$, however vacuum fields can be related to the Casimir forces and other associated effects. These definitions hold for both, for the free and for the attached fields.

Here we have defined $\mathbf{E}(\mathbf{r}_{ph},t) = \begin{pmatrix} E_x(\mathbf{r}_{ph},t) \\ E_y(\mathbf{r}_{ph},t) \\ E_z(\mathbf{r}_{ph},t) \end{pmatrix}$ and similarly for $\mathbf{B}(\mathbf{r}_{ph},t)$ and $\mathbf{A}(\mathbf{r}_{ph},t)$, and

$\mathbf{1} = \begin{pmatrix} 1 \\ 1 \\ 1 \end{pmatrix}$. Similarly we define the adjunct photon-wave-functions

$\Psi^{\dagger}(\mathbf{r}_{ph},t) = \begin{pmatrix} \bar{\mathbf{E}}^*(\mathbf{r}_{ph},t) & \bar{\mathbf{B}}^*(\mathbf{r}_{ph},t) & \bar{\mathbf{A}}^*(\mathbf{r}_{ph},t) & \phi^*(\mathbf{r}_{ph},t) & \bar{\mathbf{1}} & \bar{\mathbf{1}} & \bar{\mathbf{1}} & 1 \end{pmatrix}$ and
$\Psi_{vac}^{\dagger}(\mathbf{r}_{ph,null},t) = \begin{pmatrix} \bar{\mathbf{E}}(\mathbf{r}_{ph,null},t) & \bar{\mathbf{B}}(\mathbf{r}_{ph,null},t) & \bar{\mathbf{A}}(\mathbf{r}_{ph,null},t) & \phi(\mathbf{r}_{ph,null},t) & \bar{\mathbf{1}} & \bar{\mathbf{1}} & \bar{\mathbf{1}} & 1 \end{pmatrix}$. Where

$\vec{\mathbf{E}}^* = \begin{pmatrix} E_x^* & E_y^* & E_z^* \end{pmatrix}$ and $\vec{\mathbf{1}} = \begin{pmatrix} 1 & 1 & 1 \end{pmatrix}$. These definitions allow us to define the adjunct wave functions $F^\dagger(\mathbf{r}_e, \mathbf{r}_{ph}, t)$.

With these definitions, the operator $\hat{H}_{ph}^{free}$ takes the explicit form

$$\hat{H}_{ph}^{free} = \begin{pmatrix} 0 & \overrightarrow{(\nabla_{r_{ph}} \times)} & 0 & 0 & 0 & 0 & 0 & 0 \\ \overrightarrow{(\nabla_{r_{ph}} \times)} & 0 & 0 & 0 & 0 & 0 & 0 & 0 \\ \vec{\vec{1}} & 0 & 0 & \overrightarrow{(\nabla_{r_{ph}})} & 0 & 0 & 0 & 0 \\ 0 & 0 & (\nabla_{r_{ph}} \cdot) & 0 & 0 & 0 & 0 & 0 \\ 0 & 0 & 0 & 0 & 0 & 0 & 0 & 0 \\ 0 & 0 & 0 & 0 & 0 & 0 & 0 & 0 \\ 0 & 0 & 0 & 0 & 0 & 0 & 0 & 0 \\ 0 & 0 & 0 & 0 & 0 & 0 & 0 & 0 \end{pmatrix}$$

where $\overrightarrow{(\nabla \times)} = \begin{pmatrix} 0 & -\partial/\partial z & \partial/\partial y \\ \partial/\partial z & 0 & -\partial/\partial x \\ -\partial/\partial y & \partial/\partial x & 0 \end{pmatrix} = \nabla \cdot \mathbf{s}$, $\overrightarrow{(\nabla)} = \begin{pmatrix} \partial/\partial x & 0 & 0 \\ 0 & \partial/\partial y & 0 \\ 0 & 0 & \partial/\partial z \end{pmatrix}$ and

$(\nabla \cdot) = \begin{pmatrix} \partial/\partial x & \partial/\partial y & \partial/\partial z \end{pmatrix}$.

The $\hat{\mathbf{A}}(\mathbf{r}_{ph})$ and $\hat{\mathbf{A}}^\dagger(\mathbf{r}_{ph})$ operators for free fields can be given explicitly by the matrices:

$$\hat{\mathbf{A}}(\mathbf{r}_{ph}) = \begin{pmatrix} 0 & 0 & 0 & 0 & 0 & 0 & 0 & 0 \\ 0 & 0 & 0 & 0 & 0 & 0 & 0 & 0 \\ 0 & 0 & 0 & 0 & 0 & 0 & 0 & 0 \\ 0 & 0 & 0 & 0 & 0 & 0 & 0 & 0 \\ 0 & 0 & 0 & 0 & 0 & 0 & 0 & 0 \\ 0 & 0 & 0 & 0 & 0 & 0 & 0 & 0 \\ 0 & 0 & \vec{\vec{1}} & 0 & 0 & 0 & 0 & 0 \\ 0 & 0 & 0 & 0 & 0 & 0 & 0 & 0 \end{pmatrix} \text{ and } \hat{\mathbf{A}}^\dagger(\mathbf{r}_{ph}) = \begin{pmatrix} 0 & 0 & 0 & 0 & 0 & 0 & 0 & 0 \\ 0 & 0 & 0 & 0 & 0 & 0 & 0 & 0 \\ 0 & 0 & 0 & 0 & 0 & 0 & \vec{\vec{1}} & 0 \\ 0 & 0 & 0 & 0 & 0 & 0 & 0 & 0 \\ 0 & 0 & 0 & 0 & 0 & 0 & 0 & 0 \\ 0 & 0 & 0 & 0 & 0 & 0 & 0 & 0 \\ 0 & 0 & 0 & 0 & 0 & 0 & 0 & 0 \\ 0 & 0 & 0 & 0 & 0 & 0 & 0 & 0 \end{pmatrix}.$$

It can be verified that the operator $\hat{\mathbf{A}}(\mathbf{r}_{ph})$ applied to the photon-wave-function $\Psi$ gives

$$\hat{\mathbf{A}}\Psi(\mathbf{r}_{ph}) = \begin{pmatrix} 0 \\ 0 \\ 0 \\ 0 \\ 0 \\ 0 \\ \mathbf{A}(\mathbf{r}_{ph}) \\ 0 \end{pmatrix}, \text{ so that } \Psi_{vac}^{\dagger}(\mathbf{r}_{ph})\hat{\mathbf{A}}\Psi(\mathbf{r}_{ph}) = \mathbf{A}(\mathbf{r}_{ph}), \text{ and } \Psi^{\dagger}(\mathbf{r}_{ph})\hat{\mathbf{A}}^{\dagger}\Psi_{vac}(\mathbf{r}_{ph}) = \mathbf{A}^{*}(\mathbf{r}_{ph}) \text{ as}$$

usual in quantum optics. The operator $\hat{\mathbf{p}} \cdot \hat{\mathbf{A}}(\mathbf{r}_{ph})$ sandwiched between two wave functions gives

$$\iint F_{vac,2}^{\dagger}(\mathbf{r}_e, \mathbf{r}_{ph}, t)\hat{\mathbf{A}} \cdot \hat{\mathbf{p}}(\mathbf{r}_{ph})F_1(\mathbf{r}_e, \mathbf{r}_{ph}, t)\delta(\mathbf{r}_e - \mathbf{r}_{ph})d\mathbf{r}_{ph}d\mathbf{r}_e = \int \mathbf{A}(\mathbf{r}_e, t) \cdot (\psi_2^{*}(\mathbf{r}_e, t)\hat{\mathbf{p}}\psi_1(\mathbf{r}_e, t))d\mathbf{r}_e$$

as required by the interaction term in the particle Hamiltonian. Also, these operators fulfill the following properties: $\hat{\mathbf{A}}(\mathbf{r}_{ph})\Psi_{vac}(\mathbf{r}_{ph,null}) = 0$, $\Psi_{vac}^{\dagger}(\mathbf{r}_{ph,null})\hat{\mathbf{A}}^{\dagger}(\mathbf{r}_{ph}) = 0$. In the case of having many photons, the total operator seen by the atomic system will be $\sum_{i=0}^{n}\hat{\mathbf{A}}_i$, where $\hat{\mathbf{A}}_i$ will act only on the $i$-photon. On any other photon will produce zero. The same applies for $\hat{\mathbf{A}}_i^{\dagger}$. Its generalization for attached fields and total fields is straightforward. Similar definitions can be given for the operators for the electric and magnetic fields respectively.

Let's define the neutral-dot operator for photons by $\Psi^{\dagger}(\mathbf{r}_{ph}, t)(\bullet_{ph}^{neut})\Psi(\mathbf{r}_{ph}, t) = 1$, in matrix form:

$$(\bullet_{ph}^{neut}) = \begin{pmatrix} 0 & 0 & 0 & 0 & 0 & 0 & 0 & 0 \\ 0 & 0 & 0 & 0 & 0 & 0 & 0 & 0 \\ 0 & 0 & 0 & 0 & 0 & 0 & 0 & 0 \\ 0 & 0 & 0 & 0 & 0 & 0 & 0 & 0 \\ 0 & 0 & 0 & 0 & 0 & 0 & 0 & 0 \\ 0 & 0 & 0 & 0 & 0 & 0 & 0 & 0 \\ 0 & 0 & 0 & 0 & 0 & 0 & 0 & 0 \\ 0 & 0 & 0 & 0 & 0 & 0 & 0 & 1 \end{pmatrix}. \text{ We see immediately that } (\bullet_{ph}^{neut})^{\dagger} = (\bullet_{ph}^{neut}) \text{ and}$$

$(\bullet_{ph}^{neut})^{\dagger}(\bullet_{ph}^{neut}) = (\bullet_{ph}^{neut})$. We define a neutral-dot operator acting on free wave functions $(\bullet_{ph}^{neut})_{free}$ and a neutral-dot operator acting on attached wave functions $(\bullet_{ph}^{neut})_{att}$. Let's define the photon energy-scalar product by $\Psi^{\dagger}(\mathbf{r}_{ph}, t)(\bullet_{ph})\Psi(\mathbf{r}_{ph}, t) = E^2 + B^2$, e.g. the electromagnetic energy density, and the photon cross product operator $(\vec{\times}_{ph})$ by $\Psi^{\dagger}(\mathbf{r}_{ph},t)(\vec{\times}_{ph})\Psi(\mathbf{r}_{ph},t) = \vec{E}^* \times \vec{B}$, this is the cross product between the electric and magnetic fields, proportional to the Poynting vector. In both cases the fields are the total fields, free plus attached. Explicitly, they are written as:

$$(\bullet_{ph}) = \left[\hat{\mathbf{E}}^{\dagger}_{free}\left(\bullet^{neut}_{ph}\right)^{\dagger}_{att} + \hat{\mathbf{E}}^{\dagger}_{att}\left(\bullet^{neut}_{ph}\right)^{\dagger}_{free}\right]\left[\hat{\mathbf{E}}_{free}\left(\bullet^{neut}_{ph}\right)_{att} + \hat{\mathbf{E}}_{att}\left(\bullet^{neut}_{ph}\right)_{free}\right] +$$
$$+ \left[\hat{\mathbf{B}}^{\dagger}_{free}\left(\bullet^{neut}_{ph}\right)^{\dagger}_{att} + \hat{\mathbf{B}}^{\dagger}_{att}\left(\bullet^{neut}_{ph}\right)^{\dagger}_{free}\right]\left[\hat{\mathbf{B}}_{free}\left(\bullet^{neut}_{ph}\right)_{att} + \hat{\mathbf{B}}_{att}\left(\bullet^{neut}_{ph}\right)_{free}\right]$$

and

$$(\vec{\times}_{ph}) = \left[\hat{\mathbf{E}}^{\dagger}_{free}\left(\bullet^{neut}_{ph}\right)^{\dagger}_{att} + \hat{\mathbf{E}}^{\dagger}_{att}\left(\bullet^{neut}_{ph}\right)^{\dagger}_{free}\right]\left(\vec{\times}^{neut}_{ph}\right)\left[\hat{\mathbf{B}}_{free}\left(\bullet^{neut}_{ph}\right)_{att} + \hat{\mathbf{B}}_{att}\left(\bullet^{neut}_{ph}\right)_{free}\right]$$

where the operator $\left(\vec{\times}^{neut}_{ph}\right)$ is given by $\left(\vec{\times}^{neut}_{ph}\right) = \begin{pmatrix} 0 & \mathbf{s} & 0 & 0 & 0 & 0 & 0 & 0 \\ 0 & 0 & 0 & 0 & 0 & 0 & 0 & 0 \\ 0 & 0 & 0 & 0 & 0 & 0 & 0 & 0 \\ 0 & 0 & 0 & 0 & 0 & 0 & 0 & 0 \\ 0 & 0 & 0 & 0 & 0 & 0 & 0 & 0 \\ 0 & 0 & 0 & 0 & 0 & 0 & 0 & 0 \\ 0 & 0 & 0 & 0 & 0 & 0 & 0 & 0 \\ 0 & 0 & 0 & 0 & 0 & 0 & 0 & 0 \end{pmatrix}$, with

$\mathbf{s} = \begin{pmatrix} 0 & 0 & 0 \\ 0 & 0 & -1 \\ 0 & 1 & 0 \end{pmatrix}\hat{\mathbf{i}} + \begin{pmatrix} 0 & 0 & 1 \\ 0 & 0 & 0 \\ -1 & 0 & 0 \end{pmatrix}\hat{\mathbf{j}} + \begin{pmatrix} 0 & -1 & 0 \\ 1 & 0 & 0 \\ 0 & 0 & 0 \end{pmatrix}\hat{\mathbf{k}}$. We remark that the result of $\Psi^{\dagger}(\mathbf{r}_{ph},t)(\vec{\times}_{ph})\Psi(\mathbf{r}_{ph},t)$ is a vector in three-dimensional space.

We can define the norm-scalar product between two photon-wave-functions as $\langle\Psi|\Phi\rangle = \int \Psi^{free\dagger}(\mathbf{r}_{ph},t)(\bullet_{ph})\Phi^{free}(\mathbf{r}_{ph},t)d\mathbf{r}_{ph}$ and the norm of the function is given by $\langle\Psi|\Psi\rangle = \int \Psi^{free\dagger}(\mathbf{r}_{ph},t)(\bullet_{ph})\Psi^{free}(\mathbf{r}_{ph},t)d\mathbf{r}_{ph}$. Because we define the norm as given exclusively by the free fields, this norm is finite and allows us to work inside the Hilbert space with the photon-wave-functions.

In the case when an incident radiation field is present, the emission from the atomic dipole may present two components: one, which has the same frequency and is coherent with the incident beam, and a second, which has a different frequency and therefore is incoherent with the incident beam. The photons belonging to the coherent component are indistinguishable from the photons from the incident beam. Therefore, they must have the same coordinate as the photons from the incident beam. On the other hand, the incoherent photons are in principle distinguishable from the photons from the incident beam; therefore, must have a different coordinate, not present in the incident beam.

Accordingly, the source equation can be separated in two, depending on the coherence properties of the source. For the coherent emission we have:

$$i\hbar \frac{\partial \Psi_{coh}^{scat}(\mathbf{r}_{ph_i}, t)}{\partial t} = \hat{H}_{ph_i} \Psi_{coh}^{scat}(\mathbf{r}_{ph_i}, t) + \hat{\mathbf{j}}^{coh}(\mathbf{r}_{ph_i}, t)$$

where $\vec{r}_{ph_i}$ is the coordinate of a photon present in the incident beam, for the current density component at the same frequency as the incident field. This coherent scattering is the responsible for refraction and reflection of optical beams as well as the diffraction and extinction of x-rays in crystals. On the other hand, for the incoherent emission we have

$$i\hbar \frac{\partial \Psi_{inc}^{scat}(\mathbf{r}_{ph_s}, t)}{\partial t} = \hat{H}_{ph_s} \Psi_{inc}^{scat}(\mathbf{r}_{ph_s}, t) + \hat{\mathbf{j}}^{inc}(\mathbf{r}_{ph_s}, t)$$

where $\mathbf{r}_{ph_s}$ is a coordinate not present in the incident beam, for the current density component at a different frequency than the incident beam. The incoherent current will be the source for new photons created in the scattering process, absent from the original incident beam. In the presence of many photons we have:

$$\hat{\mathbf{j}}_{rad} = \sum_s \hat{\mathbf{j}}(\mathbf{r}_{ph_s}) = \sum_s \left( \hat{\mathbf{j}}^{coh}(\mathbf{r}_{ph_s}) + \hat{\mathbf{j}}^{inc}(\mathbf{r}_{ph_s}) \right) \text{ where } \hat{\mathbf{j}}^{coh}(\mathbf{r}_{ph_s}) = \hat{\mathbf{j}}(\mathbf{r}_{ph_s}) \delta_{\omega,\omega_0} \sum_i \delta_{is} \text{ and }$$

$$\hat{\mathbf{j}}^{inc}(\mathbf{r}_{ph_s}) = \hat{\mathbf{j}}(\mathbf{r}_{ph_s})(1 - \delta_{\omega,\omega_0})\left(1 - \sum_i \delta_{is}\right)$$ if we assume the system is contained in a finite volume and one can have a finite number of different frequencies. where $\delta_{\omega,\omega_0}$ is the delta from Kroenecker, which is equal to one if the scattered frequency equals the incident frequency, and equal to zero otherwise. In the case of an incident beam with finite width $\Delta\omega$, the $\delta_{\omega,\omega_0}$ can be replaced with the $\delta_{\omega,\omega_0,\Delta\omega}$, which signifies those frequencies already present in the spectrum of the incident beam. $\delta_{is}$ is the delta from Kroenecker, which is equal to one if the scattered photon is one of the incident photons, and equal to zero if the scattered photon is newly created, namely was not present in the incident beam.

One can assume that the photon mass is the parameter that can be used to define the distinguishability of the photon. If radiation changes frequency because of scattering, this will be reflected in its associated photon mass $m = \hbar\nu/c^2$. Therefore, one has to change the coordinates describing the scattered photon: it is not the same photon it was incident on the system, because its mass is different now.

The current density operator is given in matricial form by

$$\hat{\mathbf{j}} = \begin{pmatrix} 0 & 0 & 0 & 0 & \vec{j}_E & 0 & 0 & 0 \\ 0 & 0 & 0 & 0 & 0 & \vec{j}_B & 0 & 0 \\ 0 & 0 & 0 & 0 & 0 & 0 & \vec{j}_A & 0 \\ 0 & 0 & 0 & 0 & 0 & 0 & 0 & 0 \\ 0 & 0 & 0 & 0 & 0 & 0 & 0 & 0 \\ 0 & 0 & 0 & 0 & 0 & 0 & 0 & 0 \\ 0 & 0 & 0 & 0 & 0 & 0 & 0 & 0 \\ 0 & 0 & 0 & 0 & 0 & 0 & 0 & 0 \end{pmatrix}, \text{ so that } \hat{\mathbf{j}}_{rad} \Psi(\mathbf{r}_{ph}) = \begin{pmatrix} \mathbf{j}_E \\ \mathbf{j}_B \\ \mathbf{j}_A \\ 0 \\ 0 \\ 0 \\ 0 \\ 0 \end{pmatrix} \text{ provides precisely the}$$

current densities needed for every field in the inhomogeneous Maxwell's equations. For charged point particles we have:

$$\mathbf{j}_{E,free}(\mathbf{r}_{ph}, t) = q_e \mathbf{v}_e \delta_{free}(\mathbf{r}_{ph} - \mathbf{r}_e)$$
$$\mathbf{j}_{E,att}(\mathbf{r}_{ph}, t) = q_e \mathbf{v}_e \delta_{att}(\mathbf{r}_{ph} - \mathbf{r}_e)$$
$$\mathbf{j}_B(\mathbf{r}_{ph}, t) = 0$$
$$\vec{\mathbf{j}}_A(\mathbf{r}_{ph}, t) = 0$$

In this section, we have completed our first task, namely the presentation of a first quantization formalism that will allow us to postulate guiding conditions for the material particles including photons, and to obtain a realistic description for the atom-radiation interaction. Before going into the second task, we consider in the following Section the generalization to the case of many particles.

## V. GENERAL CASE, MANY PARTICLES

A multi-particle multi-photon, nonlinear generalization for the Hamiltonian can be the following:

$$\hat{H} = \sum_i \left( \tfrac{1}{2} m_i \nabla_{r_i}^2 + \hat{V}(\mathbf{r}_i) \right) + \sum_j \left( \hat{H}_{ph\ j}^{free} + \hat{H}_{ph\ j}^{free} \right) +$$

$$+ \sum_{\substack{n=0 \\ i,j}}^{n'} \left\{ q \nabla_{r_{ii}} \cdot \{\hat{\mathbf{A}}_j^\dagger + \hat{\mathbf{A}}_j\} + \hat{\mathbf{j}}_{rad\ ij} \right\} \delta(\mathbf{r}_{phj} - \mathbf{r}_i) \}^n$$

acting on the functions

$$F(\mathbf{r}_{e_1}, \ldots, \mathbf{r}_{e_i}, \ldots, \mathbf{r}_{e_n}, \mathbf{r}_{ph_1}, \ldots, \mathbf{r}_{ph_j}, \ldots, \mathbf{r}_{ph_{n'}}) =$$

$$= \sum_{i,j,k=0}^{n,n'} c_{ijk} \psi_1 \cdots \psi_i \Psi(\mathbf{r}_{ph_1})_1 \cdots \Psi(\mathbf{r}_{ph_j})_j \Psi_{att}(\mathbf{r}_{ph_j})_k$$

where the $i$-sum runs over the different electron's coordinates, and the $j$-sum runs over the different photon coordinates. This equation describes those cases where more than

- 24 -

one electron participate actively and coherently in the emission or absorption process, and on those cases of multi-photon emission or absorption, common in nonlinear optics.

For non-interacting, free particles the wave function must be properly symmetrized and antisymmetrized, following the postulates of indistinguishability of identical particles of quantum mechanics. In this way, the properly normalized wave function for bosons reads [120, 121]

$$F = \frac{\sum P(\psi_1(\mathbf{r}_\alpha)\psi_2(\mathbf{r}_\beta)\cdots\psi_N(\mathbf{r}_\upsilon))}{\sqrt{N!n_1!n_2!\cdots}} \tag{10}$$

where $\alpha, \beta, \cdots, \upsilon$ are the different particles composing the system, 1, 2, …, N are the states where these particles can be found, the sum is performed over all possible permutations of the particle coordinates, and $n_1, n_2, \cdots$ are the number of particles in state 1, 2, …; finally ! means the factorial as customary. We remark that this selection for the normalization also provides the right expressions for absorption and spontaneous and stimulated emission in the case of the interaction of photons with matter [122]. We should remember that in the case of photons, indistinguishability is assumed only for photons having the same frequency, this is for photons inside the same energy shell [123]. This is consistent with the consideration of the photon mass as an observable able to distinguish between photons of different frequencies. Finally, we mention that in our model, we can provide an alternative interpretation for the symmetrized wave function given in eq. (10). Instead of recalling the principle of indistinguishability of identical particles, we can equally well say that this expression is given as an initial condition for the wave function generating entangled states with strong non-local correlations. In this case, the Bose-Einstein distribution would not be a consequence of the indistinguishability of the particles, but of the correlations existing among them [124].

In the case of interacting particles, one cannot apply naively the symmetrization postulates as seen from the following considerations. The time derivative for any quantum mechanical operator is given by the commutator of this operator with the Hamiltonian. Integration of that equation provides the time-dependence of that operator. In the specific case of the creation and destruction operators for the electromagnetic field, this time dependence will be a function of free space radiation operators and, in the presence of interaction with atomic systems, of atomic operators as well. The atomic operators are in general not boson operators. Therefore, the radiation operators after interaction with matter have no more pure bosonic commutation properties. They present a mixture of bosonic, fermionic and spin properties. Because of that, the photon-wave-functions will not be strictly symmetrical. In general, the type of wave function will depend on the process of creation of the photons and the indistinguishability of the photon paths [125].

We mention here, that one has the freedom, depending on the geometry and constraints of the problem, to choose as independent variables, any linear combination of electron and photon coordinates  This will provide multiple possibilities for entanglement between electron among themselves, between photons among themselves, and between electrons and photons.   The possibility for the generation of entanglement differentiates our approach from semiclassical models, which provide no way for entanglement with photons.

For isolated atoms, spherical symmetry is the most appropriate, and the techniques developed in nuclear physics [126] and atomic physics [127] (for example the hyperspherical model) may be applied to find the solutions to the generalized wave equation, using the appropriate multi-particle functions for both, electrons and photons.  For laser beams impinging on nonlinear crystals, the use of cylindrical geometry might be more useful, and the wave functions might depend on the sum and difference of photon positions, as is the case for the biphoton [128].

If we have more than one atom irradiated simultaneously by the same light beam, all atoms may share the absorbed photon, and may start oscillating coherently in a collective fashion.  In this case, the photon-wave-function may have sources on all atoms, generating 'induced' emission with directional properties, as in the case of an array of emitting antennas.  This is the case of amplifying laser media and crystals, with refraction and reflection at the interfaces.  We can say that when the distance between scattering centers is larger than a wavelength, they are distinguishable for the incident radiation, and scattered photons have independent wave functions generating incoherent scattering.  If the distance between scattering centers is shorter than a wavelength, they are indistinguishable for the incident radiation, generating coherent scattering.

## VI. ELECTRON AND PHOTON IN REAL SPACE

Here we start with the second part of our work, namely we make the assumption that the electron as a particle has real existence in space as in Bohmian's mechanics, and furthermore we extend this assumption to the photons.  We assume additionally that the total electromagnetic fields, and not the free electromagnetic fields alone, generate the guiding condition for the photons.  In this way, we can reconcile the flux of energy in the particle model with the classical Poynting vector.   We will postulate that the photon velocity is given by the speed of propagation of energy; this is the ratio between the Poynting vector and the electromagnetic energy density, as proposed by Wesley [129].  This approach assigns to the photon-particle the speed of light in vacuum, when it is far from any charged particle, but decreases from that value in the proximities of charges due to the presence of the scattered fields.

**A. Source currents in the generalized wave equation.**

In semiclassical models [13, 21], it is customary to take the source for the electromagnetic fields as the current obtained from the Schrödinger charge density, as originally proposed by Schrödinger himself [130] and showed to be incorrect by Heisenberg [131]. Instead, we propose here that the source currents are the real charged particles actually present in space. We are induced to take this position after considering the case of an ultra-relativistic electron approaching the earth from outer space. If we take the electron-wave-function squared (e.g. the electron probability density) as the source for the electromagnetic fields, it may cover a volume of perhaps galactic dimensions, generating an almost uniform density and electromagnetic fields of immaterial magnitude. Any measurement however, will find the fields centered at the position of the electron, with a strength equal to that of any particle confined to a given location on earth.

We remark however, that we keep the usual momentum $\mathbf{p}_i = \frac{\hbar}{i} \nabla_i$ in the expression for the interaction atomic energy. This means, the electron wave function will react to an incoming electromagnetic wave, even in the case when the electron is not oscillating.

### B Total fields. Guidance conditions.

From the point of view of realistic theories for the electromagnetic field, there have been historically two approaches: the first one, due to Bohm [3, 132, 133, 134], and the second proposed by de Broglie [1, 2, 135, 136]. Bohm's approach considered the existence of a 'super-potential' acting on the electromagnetic fields in all space, and not on the photon particles. De Broglie approach considered the application of the guidance formula to spin-one-particles of finite mass, described by the Proca equation, but in the limit for infinitesimally small mass. Independently of these models, Wesley [129] proposed that the photon velocity should be given by the ratio between the Poynting vector and the energy density, namely the group velocity or velocity with which energy propagates in a medium.

In a nonrelativistic approximation, we will follow Wesley [129] who takes the guiding condition for the photon [137] as $\mathbf{v}_{ph} = \frac{\mathbf{E} \times \mathbf{B}}{E^2 + B^2}$, which can be shown to provide the correct value for the velocity of propagation of the energy even in nonlinear crystals [138], and can be obtained by following the Kemmer-Duffin-Harish-Chandra formalism [139]. A relativistic generalization can be found by following the relativistic definitions of momentum and photon density [140, 141]. We can note however, that our model will provide a very good approximation to the relativistic solution unless the source particle is moving with a speed comparable with the speed of light. An important point is that the guiding condition is a function of the position of the photon and the position of the electron as well, generating a nonlocal correlation or interaction among them.

- 27 -

We define the guiding condition for the photon as a generalization of Wesley condition:
$$\mathbf{v}_{ph} = \frac{F^*(\mathbf{r}_e,\mathbf{r}_{ph},t)(\vec{\times}_{ph})F(\mathbf{r}_e,\mathbf{r}_{ph},t)}{F^*(\mathbf{r}_e,\mathbf{r}_{ph},t)(\bullet_{ph})F(\mathbf{r}_e,\mathbf{r}_{ph},t)}.$$

Where we use the two operators $(\bullet_{ph})$ and $(\vec{\times}_{ph})$ defined previously. In a sense similar to Bohmian mechanics for electrons, one can associate with this guiding condition a photon quantum potential, generated by the electromagnetic fields, and responsible for the motion of the photons. In some sense, the attached fields resemble sources of classical potentials because they are always associated to the particles generating them. However, their non-classical nature is manifested by the fact that they can interfere with the quantum fields generated by the free electromagnetic fields. We will define the absorption of the photon, when its trajectory reaches the position of the source of the field.

On the other hand, the Quantum Potential for the electron now reads:
$$Q(\mathbf{r}_e,\mathbf{r}_{ph},t) = -\frac{\hbar^2}{2m} \frac{\nabla_{r_e}^2 \left(F^*(\mathbf{r}_e,\mathbf{r}_{ph},t)(\bullet_{ph})F(\mathbf{r}_e,\mathbf{r}_{ph},t)\right)^{1/2}}{\left(F^*(\mathbf{r}_e,\mathbf{r}_{ph},t)(\bullet_{ph})F(\mathbf{r}_e,\mathbf{r}_{ph},t)\right)^{1/2}}.$$

where we make explicit use again of the operator $(\bullet_{ph})$. This equation is equivalent to the Bohmian guiding condition:
$$\mathbf{v}_e(\mathbf{r}_e,\mathbf{r}_{ph},t) = \frac{\mathbf{j}(\mathbf{r}_e,\mathbf{r}_{ph},t)}{\rho(\mathbf{r}_e,\mathbf{r}_{ph},t)}.$$

namely the local current density divided by the particle density. In the presence of electromagnetic fields the current density is given by [142]
$$\mathbf{j} = \frac{\hbar}{2mi}\left[\psi^*\nabla\psi - (\nabla\psi^*)\psi\right] - \frac{eR^2}{mc}\mathbf{A}$$

where $R^2$ is the probability density and $\mathbf{A}$ the vector potential.

We can define an average velocity, smoothing out the variations in velocity amplitude given by the local value of the wave function, and consider the average over the volume where the particle is moving:
$$\langle \mathbf{v}(\mathbf{r}_e,\mathbf{r}_{ph},t)\rangle = \int \mathbf{v}(\mathbf{r}_e,\mathbf{r}_{ph},t)\rho(\mathbf{r}_e,\mathbf{r}_{ph},t)d\mathbf{r}_e d\mathbf{r}_{ph} = \int \frac{\mathbf{j}(\mathbf{r}_e,\mathbf{r}_{ph},t)}{\rho(\mathbf{r}_e,\mathbf{r}_{ph},t)}\rho(\mathbf{r}_e,\mathbf{r}_{ph},t)d\mathbf{r}_e d\mathbf{r}_{ph} = \int \mathbf{j}(\mathbf{r}_e,\mathbf{r}_{ph},t)d\mathbf{r}_e d\mathbf{r}_{ph}.$$

In the case where the wave function can be factorized as the product of an electron-wave-function and a photon-wave-function, the integration over the photon coordinates gives just a normalization constant, and we are left with the value of the total current as given by the interaction matrix element in the Schrödinger equation. The fields, which are produced by the real currents, will follow on the average the distribution of the Schrödinger fields. In the dipole approximation, both expressions should give practically the same result.

At this point we have finished the description of our realistic model. In the following Section we will consider its application to a two level system.

# VII. APPLICATION TO A TWO LEVEL SYSTEM

In order to illustrate the theory we have just exposed, we will apply it to the particular case of one photon and a two-level hydrogen atom.

Based on the order of magnitude of the masses involved in this problem, we can divide the wave equation in two parts: one involving the proton and the electron, with its Coulomb interaction, and other involving the photon. The proton-electron part can be solved first, and treat the photon interaction as a perturbation. This gives the traditional solutions for the hydrogen atom, where the general solution can be represented as a function of the atomic or center of mass coordinate $\mathbf{r}_A = \frac{m_p \mathbf{r}_p + m_e \mathbf{r}_e}{m_p + m_e}$ and the electron-proton relative coordinate $\mathbf{r}_E = \mathbf{r}_e - \mathbf{r}_p$. This is because the Coulomb interaction depends only on the relative electron-proton coordinate, and not on them independently. Because the proton mass is much larger that the electron mass, the position of the center of mass practically coincides with the position of the proton: $\mathbf{r}_A \cong \mathbf{r}_p$. Observation of the expressions for the fields provided by eqs. (2) and (3) shows that the fields produced by point particles do not depend on the absolute value of the field position, but only on the relative position of the photon relative to the source, apart from time. This induce us to define $\mathbf{r}_{ph} = \mathbf{r}_{field} - \mathbf{r}_A$, where $\mathbf{r}_{field}$ is the absolute photon position, and $\mathbf{r}_{ph}$ the relative position of the photon respect to the center of mass.

In the absence of external fields, the whole function for an excited atomic system can be written as:

$$F(\mathbf{r}_e, \mathbf{r}_{ph}, t) = \Psi_A(\mathbf{r}_A, t) \left[ a_g(t) \psi_g(\mathbf{r}_E, t) \Psi(\mathbf{r}_{ph}, t) + a_e(t) \psi_e(\mathbf{r}_E, t) \Psi_{vac}(\mathbf{r}_{ph,null}) \right] \Psi_{att}(\mathbf{r}_{ph}, t), \quad \textbf{(11)}$$

where $\Psi_A(\mathbf{r}_A, t)$ is the wave function for the atomic coordinate, $\psi_g(\mathbf{r}_E, t)$ and $\psi_e(\mathbf{r}_E, t)$ are the solutions of the unperturbed Hamiltonian corresponding to the ground state and excited state respectively, $\Psi(\mathbf{r}_{ph}, t)$ is the field emitted by the atomic system, $\Psi_{vac}(\mathbf{r}_{ph,null})$ are the vacuum fields and finally $\Psi_{att}(\mathbf{r}_{ph}, t)$ are the attached fields. This function shows explicitly entanglement between the photon states and the electron states. The explicit analysis of the function $\Psi_A(\mathbf{r}_A, t)$ lies outside the scope of the present work. We only emphasize that it is the responsible for the recoil of the atomic part after the photon has been ejected, and in the absence of external forces, it would be of the form of a plane wave. For a previous work on center of mass and recoil in spontaneous emission, see the review by Stenholm [143] and references therein. The entanglement between the photon and atomic system by recoil allows for the localization of photons to a region of the order of the atomic size times the ratio between (atom mass) / (photon mass), for a study of this effect in relation to the Schmidt decomposition, see the work by Eberly and coworkers [144].

In our model, the center of mass coordinate has a real existence, and follows a trajectory determined by the guiding condition for the atomic wave function $\Psi_A(\mathbf{r}_A, t)$. The other particles, including the photon, follow a guiding condition provided as a function of their relative location to the center of mass, $\mathbf{r}_E$ and $\mathbf{r}_{ph}$. The total force suffered by the particles is then the sum of that provided to the relative coordinate plus the forces acting on the center of mass.

In the presence of external radiation fields, we assume the wave function to be given by
$$F(\mathbf{r}_e, \mathbf{r}_{ph}, t) = \Psi_A(\mathbf{r}_A, t)\left(a_g(t)\psi_g(\mathbf{r}_E, t) + a_e(t)\psi_e(\mathbf{r}_E, t)\right)\Psi_{rad}(\mathbf{r}_{ph}, t)\Psi_{att}(\mathbf{r}_{ph}, t)$$
as long as the atomic system is exposed to the incident fields. Here $\Psi_{rad}(\mathbf{r}_{ph}, t)$ represents the incident and scattered fields. After the incident wave has moved away from the system, this is left in the ground state if no photon has been absorbed by the atom, or it may be left in the state described by eq. (11) above in the case when a photon has been absorbed by the system.

When a photon is absorbed, the atom starts moving because of conservation of momentum. As a result, the amount of energy available for the internal oscillation decreases, and with it the frequency of the emitted radiation. This makes the emitted photon distinguishable from the incident one, and therefore must have a new coordinate in the new wave function. The scattered photon will be incoherent with the incident beam and will not suffer interference with it. This is the Compton effect.

### A. Lamb shift and spontaneous emission.

*Semiclassical models*

In traditional quantum optics, one can obtain the correct value for the Lamb shift [145], accompanied by an exponential decay of the excited state given by spontaneous emission, while preserving the norm of the state describing the complete system. In semiclassical theories, this cannot be achieved. In fact, spontaneous emission and the Lamb shift have been considered in the semiclassical works by Jaynes and collaborators [13] and by Barut and collaborators [21]. Both approaches have problems: Jaynes' work preserves normalization, but has not an exponential decay. In Barut's work, the electromagnetic fields are allowed to take on complex values, renouncing to the unitarity of the Hamiltonian [146] and precluding the constancy of the norm of the solutions. The ground state is found stable, while excited states decay towards zero. It can be verified that these solutions have not a constant norm by considering the case where the initial state consists of an equal amount of the upper and lower states. After a few lifetimes, the excited state decays, and the system is left in the ground state, with a total norm equal to half of the original one. In relation to the exponential decay of vectors in Hilbert space, see Reference [147] where the Gamow vectors are discussed.

A third semiclassical approach is given by stochastic quantum electrodynamics [148]. Here the vacuum fields are assumed to generate forces on the radiating electron, which add to the radiation reaction forces for downward transitions. For upward transitions, the vacuum fields and the radiation reaction forces cancel each other, preventing the generation of a transition that otherwise would not conserve energy.

Another approach providing the right value for the Lamb shift is given by the dispersion forces formalism, which works with the equations generated by two requirements. The first requirement is of self-consistency: the fields seen by the particle must be the same as the fields generated by the particle [149]. The second requirements is that the energy change in the system generated by the interaction between matter and radiation should be given by $\Delta E = \sum \hbar \Delta \omega_i$, where the sum has to be performed over the atomic and radiation frequency shifts. In this model, one assumes that the real and imaginary parts of the vector potential represent the components in phase and out of phase with respect to the electron motion, as it is customary in any classical calculation using complex notation for electric fields. There is no way to generate an exponential decay. Possible solutions are periodical or steady. Steady solutions can be found by diagonalization of the Hamiltonian [150], which is always possible, independently from the phase between the velocity and the vector potential [151], and therefore also in the presence of self-fields.

We note that our model can clarify and justify the results of the dispersion forces formalism. In fact, a steady state is possible also under the consideration of radiation reaction fields, because the quantum potential cancels the forces that are out of phase with the dipole moment, preventing them from exerting any work on the particle. We notice also that the Schrödinger-Maxwell equation without interaction splits in two parts corresponding to the Schrödinger and the Maxwell equations respectively. In the absence of interaction, the total energy is given by $E = \sum \hbar \omega_i$; with interaction the change in energy has to be the sum in the changes of frequencies: $\Delta E = \sum \hbar \Delta \omega_i$, providing the second equation required by the theory of dispersion forces.

*Our model*

In this work, we propose a new model for spontaneous emission based on the entanglement between the photon and atomic wave functions. Because of that entanglement, the atom decays to the ground state and the electron stops oscillating when the photon leaves the atom. Taking a constant probability per unit time for photon emission, one obtains an exponential distribution of lengths for the photon-wave-functions, which in turn generates a Lorentzian Fourier distribution in frequency space. This model can be generalized to include collisions and other perturbations [152].

Explicitly, we are going to consider an excited state with no photons, and a ground state with one photon. Neglecting the center of mass coordinate, and taking the proton to be at rest, the wave function can be written as:

$$F(\mathbf{r}_e,\mathbf{r}_{ph},t) = \left(a_e(t)\psi_e(\mathbf{r}_e,t)\Psi_{vac}(\mathbf{r}_{ph,null}) + a_g(t)\Psi(\mathbf{r}_{ph},t)\psi_g(\mathbf{r}_e,t)\right)\Psi^{att}(\mathbf{r}_{ph}) \;. \qquad (12)$$

The generalized wave equation in the Coulomb gauge can be split into several parts as follows:

$$\frac{\partial \psi_e}{\partial t} = \nabla_{r_e}^2 \psi_e + V(r_e)\psi_e$$

$$\frac{\partial \psi_g}{\partial t} = \nabla_{r_e}^2 \psi_g + V(r_e)\psi_g$$

which are the unperturbed Schrödinger equations for the excited and ground states. In the Coulomb gauge, the photon-wave-functions $\Psi(\mathbf{r}_{ph},t)$ and $\Psi_{vac}(\mathbf{r}_{ph,null})$ give rise to equations of the form

$$\frac{\partial}{\partial t}\begin{pmatrix}\mathbf{E}\\ \mathbf{B}\\ \mathbf{A}\end{pmatrix} = \begin{pmatrix}\nabla \times \mathbf{B}\\ \nabla \times \mathbf{E}\\ \mathbf{E}\end{pmatrix} + \begin{pmatrix}\mathbf{j}_E\\ 0\\ 0\end{pmatrix}$$

which are just the usual Maxwell equations. The current densities are given by the real electron motion. Due to the large ratio between the atomic radius and the photon wavelength, and assuming that the electron speed is much less than the speed of light, we can use the dipole approximation in order to find the solution to the Maxwell equation. Assuming real photon fields, the solution to the photon equation gives approximately [153]

$$\Psi(\mathbf{r}_{ph},t) = \Psi^0(\mathbf{r}_{ph})\Theta(\mathbf{r}_{ph}-ct)e^{-i\omega_0 t} + cc,$$

where $\Psi^0(\mathbf{r}_{ph})$ is a function depending only on the photon coordinate and $\Theta(\mathbf{r}_{ph}-ct)$ preserves causality.

At this point, we are ready to show that the 'quantum jumps' associated with the emission of radiation have their origin in the atom-photon entanglement, and in the existence of material particles and photons. The electron quantum potential is given by:

$$Q(\mathbf{r}_e,\mathbf{r}_{ph},t) = -\frac{\hbar^2}{2m}\frac{\nabla_{r_e}^2\left[F^*(\mathbf{r}_e,\mathbf{r}_{ph},t)(\bullet_{ph})F(\mathbf{r}_e,\mathbf{r}_{ph},t)\right]^{1/2}}{\left[F^*(\mathbf{r}_e,\mathbf{r}_{ph},t)(\bullet_{ph})F(\mathbf{r}_e,\mathbf{r}_{ph},t)\right]^{1/2}},$$

with the wave function $F(\mathbf{r}_e,\mathbf{r}_{ph},t)$ given by eq. (12), and the definition of the operator $(\bullet_{ph})$, it takes the form

$$Q(\mathbf{r}_e,\mathbf{r}_{ph},t) = -\frac{\hbar^2}{2m}\frac{\nabla_{r_e}^2\left[\{a_e\psi_e G^{att} + a_g\psi_g(G^{free}+G^{att})\}^* \circ \{a_e\psi_e G^{att} + a_g\psi_g(G^{free}+G^{att})\}\right]^{1/2}}{\left[\{a_e\psi_e G^{att} + a_g\psi_g(G^{free}+G^{att})\}^* \circ \{a_e\psi_e G^{att} + a_g\psi_g(G^{free}+G^{att})\}\right]^{1/2}}$$

(13)

where we have defined $G^{att} = E^{att} + B^{att}$ and $G^{free} = E^{free} + B^{free}$, and where the product ∘ represents the usual dot product between vectors in space, but removing terms of the form $E \cdot B$, retaining only terms of the form $E^2$ and $B^2$. It is important to realize that the attached fields are shared by the ground state and the excited state. In addition, the attached fields are much larger than the free fields at a distance closer than one wavelength, while the free fields are much larger at distance larger than one wavelength, namely:

$G^{att} \gg G^{free}$     if    $r < \lambda$

$G^{att} \ll G^{free}$     if    $r > \lambda$,

with $\lambda$ the radiation wavelength. As a result, when the photon is close to the atom, and the attached fields are dominating, eq. (13) reduces to

$$Q(\mathbf{r}_e, \mathbf{r}_{ph}, t) = -\frac{\hbar^2}{2m} \frac{\nabla_{r_e}^2 \left[ \{a_e \psi_e + a_g \psi_g \} G^{att*} \circ \{a_e \psi_e + a_g \psi_g \} G^{att} \right]^{1/2}}{\left[ \{a_e \psi_e + a_g \psi_g \} G^{att*} \circ \{a_e \psi_e + a_g \psi_g \} G^{att} \right]^{1/2}}$$

where the attached fields can be simplified to give

$$Q(\mathbf{r}_e, \mathbf{r}_{ph}, t) = -\frac{\hbar^2}{2m} \frac{\nabla_{r_e}^2 \left[ \{a_e \psi_e + a_g \psi_g \}^* \{a_e \psi_e + a_g \psi_g \} \right]^{1/2}}{\left[ \{a_e \psi_e + a_g \psi_g \}^* \{a_e \psi_e + a_g \psi_g \} \right]^{1/2}}.$$

The photon coordinate has disappeared from the quantum potential. Consequently, the electron motion is determined by $\mathbf{v} = \dfrac{(a_e \psi_e + a_g \psi_g)^* \vec{\nabla} (a_e \psi_e + a_g \psi_g)}{(a_e \psi_e + a_g \psi_g)^* (a_e \psi_e + a_g \psi_g)}$, and the average velocity is given by $\langle \mathbf{v} \rangle = a_g^* a_e \int \psi_g^* \nabla \psi_e \, d^3 r_e + cc$, the same result as obtained by semiclassical theories.

On the other hand, once the photon has moved at a distance larger than a wavelength, the free fields predominate over the attached fields. As a result, eq. (13) reduces to:

$$Q(\mathbf{r}_e, \mathbf{r}_{ph}, t) \cong -\frac{\hbar^2}{2m} \frac{\nabla_{r_e}^2 \left( (a_g \psi_g)^* (a_g \psi_g) \right)^{1/2}}{\left( (a_g \psi_g)^* (a_g \psi_g) \right)^{1/2}}$$

which is the potential generated by the ground state alone and consequently, the average velocity is $\langle \mathbf{v} \rangle = 0$ as shown in Bohmian mechanics [154]. The passage of the photon from a region dominated by attached fields to a region dominated by free fields (or vice versa) takes place over a period of the order of the inverse of the optical frequency involved in the transition. This passage generates an abrupt change in the quantum potential seen by the electron that can be identified with the experimentally observed quantum jumps. As soon as the photon leaves the atom and reaches a distance larger than a wavelength, the electron 'sees' only the ground state function and stops its motion, and the atom remains in its ground state irreversibly. The electromagnetic wave has no more oscillating sources and vanishes soon behind the photon. This model explains the temporal behavior in experiments performed by Clauser [155], ant the antibunching effect [99].

As previously noted, our model has similitude with the original Slater model [5, 8], where the atom was assumed to emit 'ghost' or virtual fields while in the excited state. Those ghost fields were generated by virtual oscillators and propagated in space as the classical electromagnetic fields. As soon as the quantum jump occurred, and the photon was emitted, the atom stopped emitting those fields. The original Slater model was eventually abandoned in favor of the so-called Bohr-Kramers-Slater model [156], where the photon particle was removed, surviving only the notion of statistical processes induced by virtual oscillators and virtual fields. This was done before the introduction of wave functions and the Schrödinger equation in quantum mechanics.

The equations for the coefficients read:
$$\frac{\partial a_e}{\partial t} = a_g \int \mathbf{A}(\mathbf{r}_e) \cdot \left( \psi_e^*(\mathbf{r}_e) \nabla_{r_e} \psi_g(\mathbf{r}_e) \right) d\mathbf{r}_e(\mathbf{r}_e) \cong a_g \mathbf{A}_{self}(\mathbf{r}=0,t) \cdot \mathbf{p}_{eg}$$
and
$$\frac{\partial a_g}{\partial t} = a_e \int \mathbf{A}(\mathbf{r}_e) \cdot \left( \psi_g^*(\mathbf{r}_e) \nabla_{r_e} \psi_e(\mathbf{r}_e) \right) d\mathbf{r}_e(\mathbf{r}_e) \cong a_e \mathbf{A}_{self}(\mathbf{r}=0,t) \cdot \mathbf{p}_{ge}$$
where we have assumed that the photon-wave-functions are normalized to unity. In the absence of external fields, the transverse vector potential due to the self-field at the position of the atom has been calculated both, semi classically [157] and fully quantum mechanically to be (in complex notation) [158, 159]
$$\mathbf{A}_{self}(\mathbf{r}=0,t) = \frac{q\mathbf{v}(t)\omega_0}{6\pi\varepsilon_0 c^3} \left[ i + \frac{2}{\pi} \ln\left(\frac{mc^2}{\hbar\omega_0}\right) \right]$$
where $\omega_0 = \frac{E_e - E_g}{\hbar}$, and the real and imaginary parts represent the in phase and out of phase components of the self field respectively. It is interesting to remark that the magnitude of the self-fields is the same as the magnitude of dipole fields at a distance of about a wavelength from the origin.

Similarly to the dispersion fields model, we look for steady solutions to these equations that can be obtained by diagonalization of the Hamiltonian. In those solutions, the effect of the self-fields is not to induce the decay of the oscillations as in the Jayne's and Barut's models, but to provide the right phase difference between the atomic wave functions and the electromagnetic wave. The choice for steady solutions allows for a constant probability per unit time for photon emission, and therefore for a Lorentzian Fourier distribution in frequency space, as previously mentioned.

When one considers not just a two level system, but allows for coupling between all atomic levels, this procedure provides the correct value for the Lamb shift. One can state

- 34 -

that a self-sustained state of oscillation is achieved, as equilibrium between the quantum forces and the radiation reaction forces is realized. Because of the coupling through all possible transitions, the electron motion acquires an additional drifting, filling up the space surrounding the nucleus with a probability proportional to the square of the magnitude of the wave function, in a similar way as proposed by stochastic electrodynamics. Indeed, those random motions generate electromagnetic fields very similar to the customary vacuum fields described in stochastic electrodynamics.

## B. Scattering, absorption, dispersion and resonance fluorescence

The processes of photon scattering and resonance fluorescence are described in quantum optics by the Bloch equations for the density matrix [160 60]. In our model, we don't need to add ad hoc decay terms, because spontaneous emission is described by the combined dynamics of the wave functions and the particles of the system. It is well known that when the transitions produced by the quantum jumps are incorporated into the equations, we recover the Bloch equations [104, 105, 106].

The classical equations of motion for a dipole in the presence of an external field describe two independent degrees of freedom: the amplitude and velocity of the dipole. On the other hand the Bloch equations are describing a physical system with three degrees of freedom [161]: two representing the electric dipole as in classical electrodynamics, plus another one, representing the wave function through the atomic inversion. Our model explains automatically this third degree of freedom because our system is composed not only by the particle, but also by the wave function, which follows the generalized wave equation. The wave function interacts with the particle through the quantum potential. In the case of resonance fluorescence, the solution to the Bloch equations describes a time-dependent phase difference between the oscillating dipole and the incident field. From a purely classical point of view, there is no obvious way to explain this behavior. From our model, we can verify that the phase of the dipole is modified periodically by the quantum potential in just the right way to generate the Rabi oscillations.

### *1. One photon*

Let's consider an incident electromagnetic wave for a single photon, on a two-level atom from the standpoint of our model. As shown by the dispersion forces model and the semiclassical Floquet theory as described by Shirley [162], due to the atom-radiation interaction, eigenstates of the atom and radiation system are different from the eigenstates they had when they were far apart. As a result, the system enters a state, which is not an eigenstate of the unperturbed Hamiltonian. It is of the form
$$F = \left(a\psi_{\text{ground}}^{\text{atom}} + b\psi_{\text{excited}}^{\text{atom}}\right)\Psi^{rad}\Psi_{att}$$
where $\psi_{\text{ground}}^{\text{atom}}$ and $\psi_{\text{excited}}^{\text{atom}}$ are the eigenstates of the atomic system in the absence of the incident radiation and similarly, the radiation field is the sum of the incident and the scattered fields $\Psi^{rad} = \Psi^{inc} + \Psi^{scatt}$. The coefficients $a$ and $b$ are determined by the condition that the frequency of the system remains equal to the frequency of the incident photon in vacuum.

This model is not equivalent to the dressed states model [163] because of the presence of the second term $b\psi_{excited}^{atom}\Psi^{rad}$. In fact, the dressed state corresponding to this situation is given by [164] $F = a|g,1\rangle + b|e,0\rangle$, where the atomic excited state appears only in the absence of radiation. From the very beginning, we are considering the system in a dynamical state, and as a result, we don't need to postulate that the electromagnetic field is in a Glauber coherent state, as one is forced to do in order to derive the Bloch equations in the dressed atom approach [163]. In fact, we are considering here that the field state is the equivalent to a one-photon-state in quantum optics.

The real existence of these mixed states has been signaled by some authors [165]. The incorporation in the system of terms proportional to $\psi_{excited}^{atom}\Psi^{rad}$ plays also a central part in the polaron-exciton description of the index of refraction in crystals. In fact, in Hopfield's model for the index of refraction [166], the basic crystal states are taken to be the quantized modes of the atomic crystal, implying that even in the ground state, collective oscillations of the atomic charges exist, generating a finite dynamical atomic polarization.

We remark here that the inclusion of the term $b\psi_{excited}^{atom}\Psi^{rad}$ does not imply a violation to the law of conservation of energy. The radiation field is not equal to the field in vacuum: it includes the scattered fields and as a consequence, its energy contains decreases. This decrease in the photon energy is employed by the atomic system to include the wave function from the excited state and to stay in oscillation. A similar process occurs in the classical scattering of radiation, where part of the energy of the incident field is shared by the scattered fields and the scattering center. This effect can be more clearly seen in a crystal, where the equivalent to the photonic kinetic energy given by the $\hat{H}_{ph}(\mathbf{r}_{ph})$ term in eq. (7) diminishes as a result of the increase in wavelength produced by the index of refraction. Even in the case of high intensity fields, the atomic system can perform coherent Rabi oscillations without the need for absorbing a photon. The energy needed for the excitation of the system can be provided by the full set of photons keeping the change in energy for every individual photon very small. In this way, one can explain not only the index of refraction, but also all nonlinear coherent effects associated with the propagation of pulses in transparent materials, which are adequately explained by semiclassical theories without the need for radiation quantization [167].

It can be shown that this theory provides the same energy levels as those predicted by a full quantum electrodynamics treatment [168, 169].

If we are not concerned with the motion of the atom, we can rewrite the solution to the generalized wave equation for an atom in the presence of a single incident photon field in more detail as:

$$F(\mathbf{r}_e, \mathbf{r}_{ph}, t) = \{a_g(t)\psi_g(\mathbf{r}_e, t) + a_e(t)\psi_e(\mathbf{r}_e, t)\}\Psi^{rad}(\mathbf{r}_{ph}, t)\Psi_{att}(\mathbf{r}_{ph}, t) \ .$$

The equations for the coefficients $a_e$ and $a_g$ follow traditional perturbation theory. One has to be careful to include the self-fields from the charged particles in the definition of the total fields as in the previous case of spontaneous emission and Lamb shift. It is important also to analyze when the incident radiation field should be considered as an adiabatic or as a sudden perturbation [170, 171, 172]. In the adiabatic case, one is dealing with a static approximation [173], where the system follows that eigenstate of the energy, which evolved from the initial state. This case applies to low intensity fields or frequencies far from the atomic frequencies. It gives rise to dispersion and the index of refraction. In the sudden case, one is in the presence of a dynamical [174] evolution generating oscillations at the Rabi frequency [48]. This case applies in resonance fluorescence; in this case, the main result of the excitation is the incoherent scattering of radiation. This effect is strictly related to the distinguishability of the emitted photon [175]. In both cases, the effect of the self-fields is not to induce the decay of the oscillation, but to provide the right phase difference between the atomic wave functions and the incident wave. This phase difference in turn, will provide the probability that an incoming photon be absorbed by the atomic system and will provide the introduction of the damping constant into the solutions of the problem. Transitions between wave functions with different number of photons take place in our model, only when a photon is physically absorbed or emitted by the atomic system.

Far from resonance, the equation for the incident photon reads, on the average:

$$\frac{\partial \Psi^{rad}(\mathbf{r}_{ph}, t)}{\partial t} = \hat{H}_{ph}\Psi^{rad}(\mathbf{r}_{ph}, t) + a_e^* a_g \int \psi_g^* \nabla_e \psi_e d^3 r_e \ .$$

Without considering numerical factors of order one, the net power exchange between the incident fields and the dipole through integration over a closed surface of $\mathbf{S}_i$ is classically given by the work done on the dipole $P_{cl} \approx ev E$. Solution of the classical equation of motion for an oscillator of natural frequency $v_0$ gives for the amplitude of the steady state oscillation $x = \dfrac{Ee/m}{v_0^2 - v^2 - iv\gamma}$. Here $\gamma$ is proportional to the damping force produced by the reaction fields and is given by $\gamma = \dfrac{2}{3}\dfrac{e^2 v_0^2}{mc^3}$. At resonance, the velocity is given by $v \cong vx = \dfrac{3}{2}\dfrac{Ec^3}{ev^2}$ and the power by $P_{cl} \approx \dfrac{3}{2}\dfrac{E^2 c^3}{v^2} = \dfrac{3}{2}E^2 c\lambda^2$, which defines the cross section in resonance as roughly $\lambda^2$. These classical solutions reproduce the main features of the quantum mechanical solution including the self-fields of the electron for frequencies far from the resonance frequency and even at resonance for low beam intensities..

From the steady state solution for $x$, we can see that for photons with an energy lower than the transition energy, the phase shift is negative, while for higher energies, it is positive. Simulations of the photon paths for an incident beam on an atomic system under the dipole approximation are shown in Figures 1 and 2. The simulations show, that for a positive phase shift between the incident and scattered fields, the photons are absorbed by the dipole, some of them may perform a few orbits before reaching the dipole position. On the contrary, for a negative phase shift, the photons are being repelled by the dipole fields. This means that for photons with an energy lower than the transition energy, the fields are repulsive, preventing the photons from reaching the atom. For higher energies, the fields are attractive, allowing the photons to be absorbed by the atom, which agree with the intuitive view that only photons with enough energy can be absorbed by the system.

In the case of resonance fluorescence, the phase difference between the oscillating dipole and the incident beam is a function of time, determining the sign of the work done by the external fields. When the Bloch state vector is moving up, the fields do work over the atom, this last one absorbing energy, and the photons are allowed to be absorbed. When the Bloch state vector is moving down, the atom is delivering energy and photons to the fields. Obviously, incoherently scattered photons may be emitted at any time.

## *2. N-photons*

Dirac showed in his 1927 paper [66] that a proper symmetrization and normalization of the photon-wave-function automatically generates the correct expressions for stimulated absorption and stimulated and spontaneous emission. If we have $n$-external incident photons, we can write the wave function as

$$F(\mathbf{r}_e, \mathbf{r}_{ph1}, \cdots, \mathbf{r}_{phn}, t) = \left(a_g \psi_g^n + a_e \psi_e^n\right) \Psi_n^{inc} \Psi_{att}^{inc} +$$
$$+ \sum_i \Psi_{n-i}^{inc} \left(a_{ei} \psi_e^{n-1} \Psi^{vac} + a_{gi} \psi_g^{n-1} \Psi_{n-i}^{inc} \left(\Psi_s^{scatt} + \Psi_i^{scatt}\right)\right) \Psi_{att,n-1}^{inc}$$

where the first term is present as long as no photon has been absorbed, and the second term appears once an incoming photon has been trapped by the atomic system. Here $\Psi_n^{inc}(\mathbf{r}_{ph2}, \cdots, \mathbf{r}_{phn}, t)$ is the wave function for $n$ incident photons, and $\psi_g^n(\mathbf{r}_e, t)$ and $\psi_e^{n-1}(\mathbf{r}_e, t)$ are the electron-wave-functions for the ground state and the excited state in the presence of $n$ and $n-1$ incident photons respectively. These wave functions differ slightly from the naked $\psi_g(\mathbf{r}_e, t)$ and $\psi_e(\mathbf{r}_e, t)$ and their energy has a shift known as the AC Stark effect. Here $\Psi_i^{scatt}(\mathbf{r}_{ph_i}, t)$ is the wave function for a photon with coordinate $\mathbf{r}_{ph_i}$, whose source will be different from zero only if the atom is oscillating at the same frequency as the incident field. This photon is coherent with the incident beam and indistinguishable from the incident photons; the coordinate $\mathbf{r}_{ph_i}$ is one of the coordinates present in the incident beam. Similarly $\Psi_s^{scatt}(\mathbf{r}_{ph_s},t)$ as the wave function for a newly created photon with coordinate $\mathbf{r}_{ph_s}$, whose source is the incoherent atomic current generated if the atom is in a state which is oscillating at a different frequency than the incident beam. This photon is incoherent with the incident beam and distinguishable from the incident photons; the coordinate $\mathbf{r}_{ph_s}$ is absent from the incident beam. $\Psi_{n-i}^{inc}$ is the normalized wave function of $n-1$ variables when the photon $i$ is missing. $\Psi_{n-1}^{inc}$ is the normalized wave function for the incident photons, when one photon has been physically absorbed by the atomic system. Its explicit form is given by:

$$\Psi_{n-1} = \sum_{i=1}^{n} \frac{\Psi_{n-i}}{n} = \sum_{i=1}^{n} \frac{\Psi(\mathbf{r}_1,\cdots,\mathbf{r}_{i-1},\mathbf{r}_{i+1},\ldots,\mathbf{r}_n,t)}{n}$$

where the sum runs over all the initially incident photons. One can verify explicitly that the $n$ terms correctly gives the single photon cross section as $1/n$ fraction of the total cross section as provided by the Bloch equations [176], and the total scattered intensity is $n$ times the intensity scattered per photon.

In order to generate the antibunching effect, we postulate that terms of the form $a_{gij}\psi_e^n \Psi^{scatt}(\vec{r}_{ph1})\Psi^{scatt}(\vec{r}_{ph2}) \sum_{i,j} \Psi_{n-i-j}^{inc}$ where we have two scattered photons and the atom in the ground state, and $\Psi^{scatt} \sum_{i} a_{ei}\psi_e^{n-2}\Psi_{n-i}^{inc}$ where we have one scattered photon and the atom in the excited state are null until the first scattered photon particle is actually emitted. This is in agreement with the theories of multiphoton emissions [177] and quantum jumps.

## VIII. RELATIVISTIC GENERALIZATION

In this Section, we consider a relativistic generalization of our model and we discuss its relationship with quantum electrodynamics or QED.

The first step to generalize the Schrödinger-Maxwell equation is to replace the Schrödinger part by the Dirac expression. The next step is to verify that the single particle wave equations show a relativistic covariance and that the equation of motion for the electron and photon particles leads to a relativistic four-current expressions. Finally, the multi-particle formalism can be recovered as in the case of relativistic QED, through a multi-time formalism.

We start by repeating the generalized wave equation

$$i\hbar \frac{\partial F(r_1, r_2, \cdots, r_n, r_{ph})}{\partial t} = \hat{H} F_{DM}(r_1, r_2, \cdots, r_n, r_{ph})$$

where now the generalized Dirac-Maxwell Hamiltonian corresponds to the addition of the Dirac operators for electrons [178] and the Maxwell operator for photons, corresponding to free fields and attached fields as discussed in the previous Sections.

$$\hat{H}_{DM} = \underbrace{\beta mc^2 + c\boldsymbol{\alpha} \cdot [\hat{\mathbf{p}} - q\mathbf{A}(\mathbf{r})] + qV(\mathbf{r})}_{Dirac} + \underbrace{\hat{H}_{ph}^{free}(\mathbf{r}_{ph}) + \hat{\mathbf{j}}^{free}(\mathbf{r}_{ph}) + \hat{H}_{ph}^{att}(\mathbf{r}_{ph}) + \hat{\mathbf{j}}^{att}(\mathbf{r}_{ph})}_{Maxwell}$$

The electron part as provided by the Dirac equation implies a four-current density given by [179]

$$j^\mu \equiv \bar{\psi} \gamma^\mu \psi = (\psi^\dagger \psi, \psi^\dagger \boldsymbol{\alpha} \psi)$$

where the $\gamma^\mu$ are the so-called relativistic Dirac matrices. Because the Dirac equation is relativistically invariant, and the four-current density corresponds to the assumed motion of the real electron particles, they are relativistic as desired. In fact, it can be shown that this current density is equivalent to the relativistic guiding condition for electrons [180].

The relativistic invariance of the photon operator $\hat{H}_{ph}(\mathbf{r}_{ph})$ was confirmed by many authors [72], and which is obvious because it is simply a formatted expression for the Maxwell equations. The division between free and attached fields, where the free fields are given by the acceleration fields, and the attached fields by the velocity fields is a relativistic partition as can be seen by writing the corresponding field-strength tensors in explicit covariant form [181]:

$$F_{free}^{\mu\nu} = \frac{e}{4\pi} \frac{1}{R^3} \left[ (x-z)^\mu \ddot{z}^\nu R - (x-z)^\nu \ddot{z}^\mu R - (x-z)^\mu \dot{z}^\nu Q + (x-z)^\nu \dot{z}^\mu Q \right]$$

and

$$F_{att}^{\mu\nu} = \frac{e}{4\pi} \frac{1}{R^3} \left[ (x-z)^\mu \dot{z}^\nu - (x-z)^\nu \dot{z}^\mu \right]$$

where

$$R = (x-z)^\sigma \dot{z}^\sigma ,$$
$$Q = (x-z)^\sigma \ddot{z}^\sigma ,$$

$x^\mu$ is the fixed field world-point and $z^\mu(s)$ is the source world-point describing the position of the moving charge as a function of the proper time $s$, and $\dot{z}^\mu(s)$ is the four-velocity of the particle.

The photon velocity is described by the zero-components of the symmetric stress tensor for electromagnetic radiation $\Theta^{ij}$, which define a relativistic four-vector equivalent to the electromagnetic energy density and the Poynting vector [182]:

- 40 -

$$\Theta^{00} = \frac{1}{8\pi}\left(E^2 + B^2\right) = u$$

$$\Theta^{0i} = \frac{1}{4\pi}(\mathbf{E}\times\mathbf{B}) = c\mathbf{g}$$

where $u$ is the electromagnetic energy density and $\mathbf{S} = c^2\mathbf{g}$ is the Poynting vector. In regions of space free of sources, the conservation law for momentum for the free fields reads:

$$0 = \partial_\alpha \Theta^{\alpha 0} = \frac{1}{c}\left(\frac{\partial u}{\partial t} + \nabla\cdot\mathbf{S}\right)$$

which allows for the definition of a photon current $\mathbf{v}_{ph} = \frac{\mathbf{S}}{u}$, as provided by our photon guiding condition. The fact that the integral over space of the energy density is not a relativistic constant (has different values for different frames of reference) is not of concern here, because in our interpretation the primary role of the fields is as providers for the particle guiding conditions, and not as a particle density. In this sense, a renormalization of the energy density and Poynting vector is allowed for different frames of reference without conspiring against the internal consistency of the theory. This fact has been also analyzed by various authors [183].

The relativistic generalization to many particles can be done analogously to the multiple-time formalism introduced by Dirac [184].

In a similar way as that follow in eq. (9) from Section IV, we can show that our approach is equivalent to a first quantization formalism for QED, where the wave functions fulfill the Heisenberg equation of motion as provided by QED, complemented with the existence of real particles following realistic trajectories in space and time. Finally, we remark that semiclassical models have been able to obtain the correct expressions for the Lamb shift in conjunction with spontaneous emission [21, 149].

## IX. ANALYSIS OF PREVIOUS CRITICS

The criticism of Oppenheimer [71] to the original theory of Landau and Peierls [69] was based on the difficulties associated with the appearance of negative energies implying the absence of a positive definite photon density. Formal solutions to those problems have been given by M. Hawton and collaborators [96]. We can illustrate the problem by considering the following wave function, which was discussed by Bohm and Hiley for bosons in general [185]:

$$\phi(x,t) = \phi_+(x)e^{-iEt} + \phi_-(x)e^{+iEt} \qquad (14)$$

allowing for both, positive and negative energy solutions. The particle density one obtains using the relativistic expression for current densities is:

- 41 -

$$j_0 = \phi_+^*(x)\phi_+(x) - \phi_-^*(x)\phi_-(x) + \phi_+^*(x)\phi_-(x)e^{2iEt} + cc ,$$

which obviously is not positive definite. In our model, we can employ a solution inspired by second quantization: we assume that the functions $\phi_+(x)e^{-iEt}$ and $\phi_-(x)e^{+iEt}$ are associated with two different particles, in this case antiparticles. In this way we have two independent particle densities $j_{0+} = \phi_+^*(x)\phi_+(x)$ and $j_{0-} = \phi_-^*(x)\phi_-(x)$, both positive definite, and the cross term doesn't exist. The total field however, as seen by other particles is still given by the sum in eq. (14) as required. This procedure can be further justified by the approach followed in this work, where states corresponding to different frequencies are distinguishable, and therefore are associated with different particle coordinates.

Two objections are generally presented against a separation between fields attached to the particle and free fields [51]. The first objection remarks that an extended field tied to a particle generates a non-local correlation between the field and the particles and should not be considered in a relativistic theory. In this work, we show that the non-localities appearing in the theory are of the same type as the non-localities generated by entanglement in traditional quantum mechanics, and therefore there is no reason to exclude them. The second objection is based on the idea that the uncertainty principle removes the concept of trajectories, and therefore also of fields localized around the position of the source  The mere existence of our model, consistent with the quantum rules, where real particles follow well defined trajectories given by the guidance conditions, shows that this objection is really artificial.

Very recently, it has been claimed that differences between the experimental predictions from Bohmian trajectories and the traditional interpretation of quantum mechanics exist [186], and that experiments have shown that the traditional predictions are the right ones [187]. The proposed experiments are based on the measurement of coincidences between two photons reaching a double slit simultaneously. According to traditional quantum mechanics, coincidences should occur independently of the position of the detectors. Calculations performed on Reference [186] find that coincidences should occur only when the two detectors are on different sides respect to the plane separating the two slits. It can be verified, however, that the alleged differences have their origin on the symmetric distribution of the initial trajectories attributed to the trajectories assumed in Reference [186]. This implies an excessive restriction absent from a correct formulation of the problem. If one removes that restriction, and finds the trajectories assuming asymmetric as well as symmetric initial conditions, the results from traditional quantum mechanics are recovered, and therefore no difference exist between both theories. It should be stressed that the inclusion of trajectories for the particles in no way modifies the results from quantum mechanics, because those trajectories by construction, are exactly those that are needed to reproduce the quantum mechanical predictions.

Finally, we mention that Bohmian mechanics has been mainly criticized because the particles cannot modify the dynamics of the fields. In the model presented in this work, on the other hand, there is a mutual interrelationship between particles and fields. The dynamic of the fields is modified by the dynamic of the particles and vice versa.

# X. CONCLUSIONS

We have shown that it is possible to understand the quantum mechanical atom-photon interaction from a realistic point of view. This is achieved through the application of a first quantized version of quantum electrodynamics and the postulation of the real existence of electrons and photons in three-dimensional space. The first quantized version of quantum electrodynamics provides a unified description for charged and photon particles, through a generalized equation, combining the Schrödinger and Maxwell equations. The solutions to this equation are functions of the electron coordinates and photon coordinates in an extended configuration space. The existence of electrons and photons in three-dimensional space allows for a deterministic understanding of quantum effects in general, and for quantum jumps in particular.

We show that a proper normalization of the photon-wave-functions is possible by the separation of the electromagnetic fields into fields attached to the particle and free fields carrying the radiated energy. This separation also allows for the natural entanglement between different states of the atomic systems and the electromagnetic fields. These correlations cannot be present in a semiclassical description of the electromagnetic field.

We were able to show the physical origin of the Lamb shift as the joint work of the quantum potential and the real electron oscillating in it. In the ground state, the quantum potential shields completely all forces, inclusive radiation reaction forces, preventing therefore the collapse of the electron towards the nucleus. In this model, the uncertainty principle becomes a consequence of the action of the quantum potential on the particle dynamics.

We can identify the similitude between the vacuum fluctuations of the radiation field and the resonant self-sustained oscillations reached between the charged particle, the quantum potential and radiation reaction forces. This interpretation opens the door for the engineering of the fluctuating radiation fields, whose source should not be sought of in infinity as usually interpreted, but at the real oscillations of the particles in the neighborhood of our system. This provides a realistic interpretation of superradiant states and the collective motion of atoms and molecules inside a crystal, even in their ground state.

Finally, we remark that the representation of this model in three-dimensional space makes it especially adequate for the simulation of quantum devices in quantum computation and quantum information systems.

## Acknowledgements

The author would like to thank Dr. Clay Stewart and Anu Bowman, for support during the development of this work. In addition, the author is thankful to Dr. Steve Israel and Ted Yachik for valuable comments on the manuscript.

**Figure captions.**

**Figure 1. Photon trajectories for an attracting dipole in resonance. The photons converge to the oscillating dipole. The angle between the dipole and the field is Π/4. Some of the photons make orbits before being absorbed. The photons are moving from left to right. The picture shows the motion over about 1 period of oscillation.**

**Figure 2. Photon trajectories for a repelling dipole in resonance. The photons are repelled by the oscillating dipole. The angle between the dipole and the field is - Π /4. The photons are moving from left to right. The picture shows the motion over about 1 period of oscillation.**

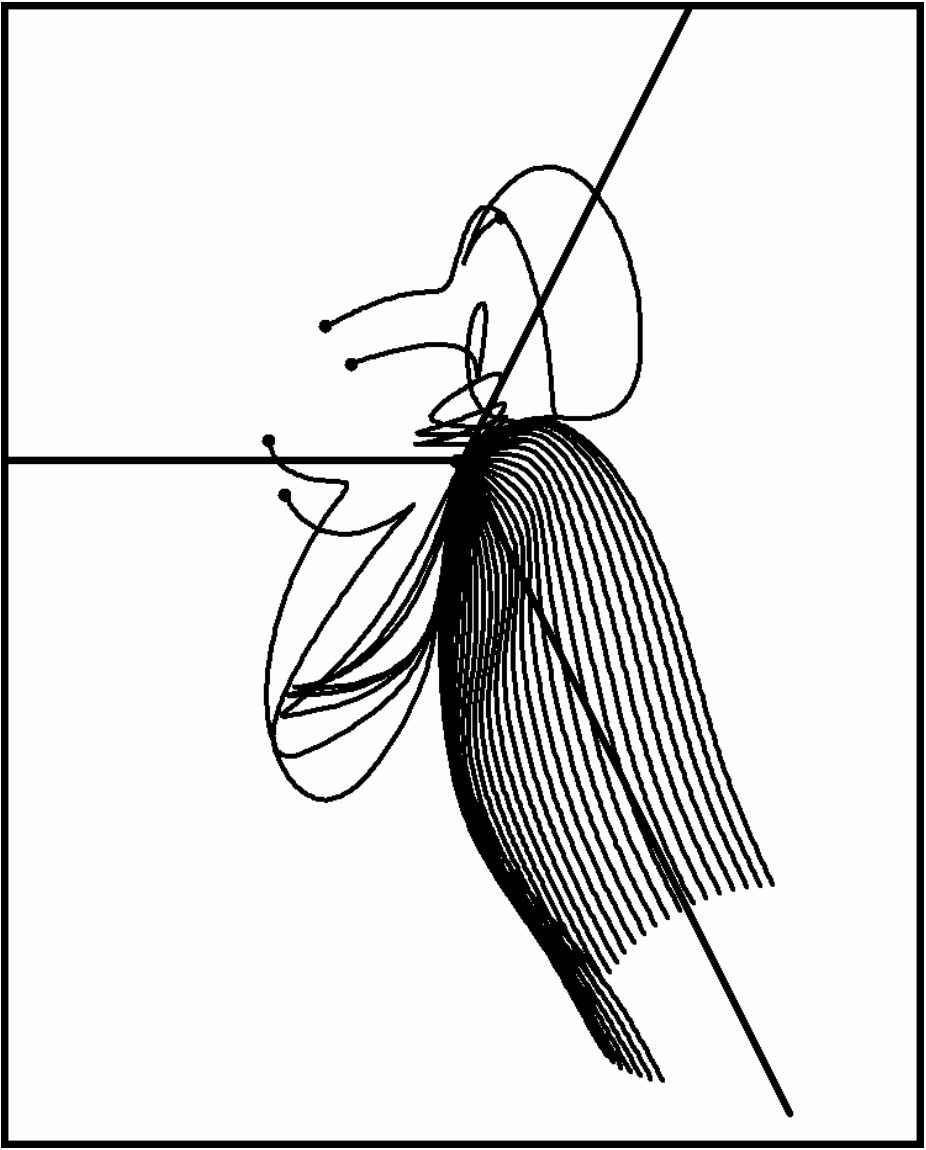

Figure 1.

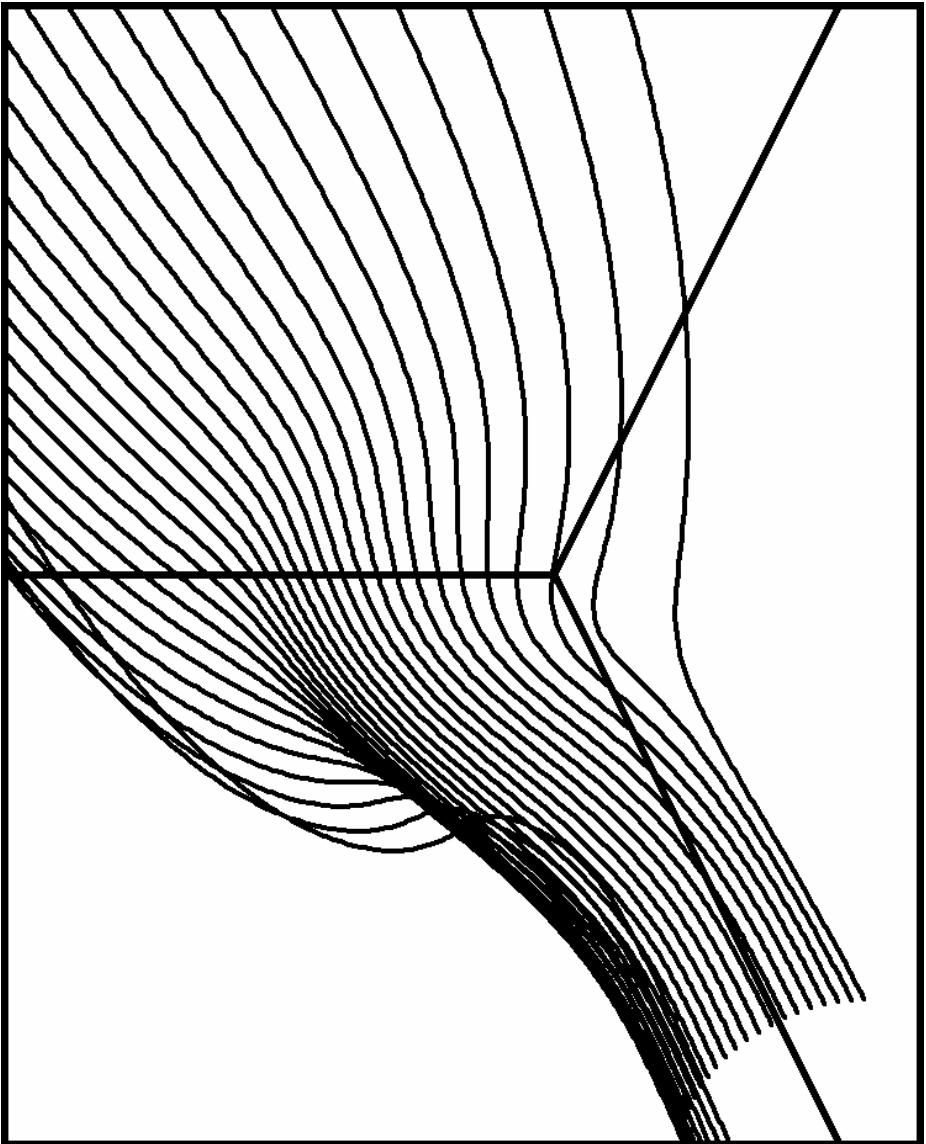

Figure 2.